\documentclass[manuscript]{aastex}
\usepackage{emulateapj5}

\received{}
\accepted{}
\shorttitle{Pulsar Beam Geometry}
\shortauthors{Everett \& Weisberg}

\begin{document}

\title{Emission Beam Geometry of Selected Pulsars Derived from Average Pulse Polarization Data}

\author{J. E. Everett\altaffilmark{1}}
\and
\author{J. M. Weisberg\altaffilmark{2}}
\affil{Department of Physics and Astronomy, Carleton College,
    Northfield, MN 55057}
\email{everett@apollo.uchicago.edu}
\email{jweisber@carleton.edu}

\altaffiltext{1}{Also Department of Astronomy and Astrophysics,
University of Chicago, 5640 S. Ellis Street, Chicago, IL, 60637. }

%\altaffiltext{2}{Email: {\it{jweisber@carleton.edu}}}

\slugcomment{To be submitted to ApJ.  Draft: September 13, 2000 jee}

\begin{abstract}
By fitting the classical Rotating Vector Model (RVM) to high quality
polarization data for selected radio pulsars, we find the inclination
of the magnetic axis to the spin axis, $\alpha$, as well as the
minimum angle between the line of sight and the magnetic axis,
$\beta$, for ten objects.  We give a full treatment of statistical
errors in the fitting process.  We also present a dictionary and
conversion table of various investigators' geometric definitions to
facilitate future comparisons. We compare our results with other RVM
fits and with empirical / geometrical (E/G) approaches, and we examine
the strengths and weaknesses of RVM fits and E/G investigations for
the determination of pulsar emission beam geometry.

Our fits to B0950+08 show that it is an orthogonal rotator with the
main and interpulse radiation emitted from opposite magnetic poles,
whereas earlier RVM fits indicated that it is an almost--aligned,
single--magnetic pole emitter.  We demonstrate that low--level
emission across a wide longitude range, when properly weighted in the
RVM fit, conclusively favors the former scenario.  B0823+26 is also an
orthogonal rotator. We find that B1929+10 emits into its wide observed
range of longitudes from portions of a single cone that is almost
aligned with the spin axis.  This result agrees with other RVM fits
but conflicts with the E/G findings of \citet{rr97}.

We determine that convergent RVM solutions can be found only for a
minority of pulsars: generally those having emission over a relatively
wide longitude range, and especially those pulsars having interpulse
emission.  In pulsar B0823+26, our preferred fit to data atall
longitudes yields a solution differing by several $\sigma$ from a fit
to the main pulse / postcursor combination alone.  For pulsar
B0950+08, separate fits to the main pulse region, the interpulse
region, and our preferred fit to almost all longitudes, converge to
results differing by several times the formal uncertainties. These
results indicate that RVM fits are easily perturbed by systematic
effects in polarized position angles, and that the formal
uncertainties significantly underestimate the actual errors.

\end{abstract}


\section{Introduction}
\begin{figure*}
\centerline{\includegraphics[width=3.5in, height=4.0in]{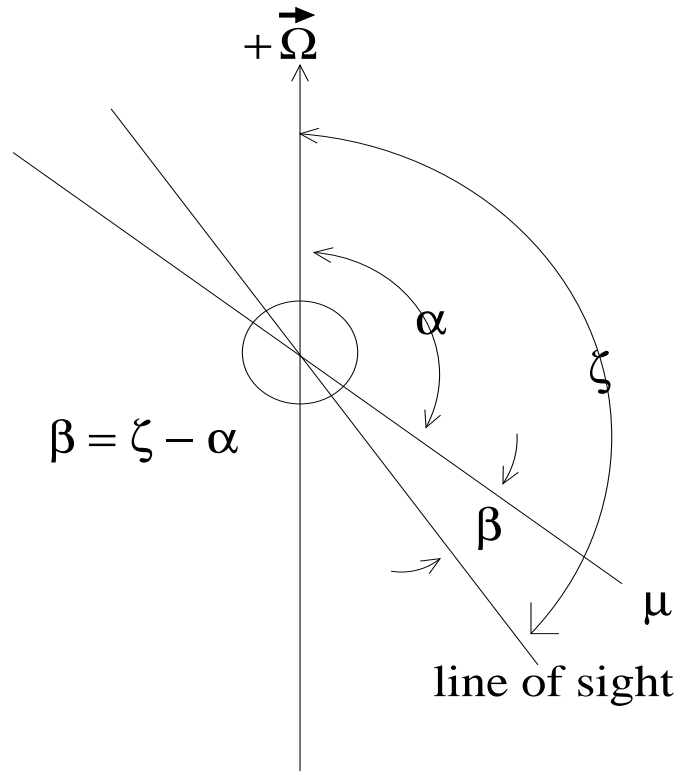}}
\caption{In the rotating vector model (RVM, see Eq. 1), the
magnetic pole colatitude $\alpha$, line of sight colatitude $\zeta$,
and line of sight impact parameter $\beta$ are defined in terms of the
positive angular velocity vector $+\vec{\Omega}$ and the observable
magnetic pole $\mu$.  These quantities are shown here in the plane
containing $\vec{\Omega}$ and $\mu$.  Note that these definitions are
the same for all possible $\alpha$; i.e., for $(0\leq\alpha\leq\pi)$
radians.
\label{fig-1}}
\end{figure*}

Determination of the geometry of radio pulsar emission is essential to
understanding emission mechanisms.  The orientation of any pulsar
reduces basically to two angles: $\alpha$, the angle between the spin
axis and the observable magnetic axis, and $\beta$, the minimum angle
between the magnetic axis and the observer's line of sight as the beam
sweeps past the observer (see Fig.~\ref{fig-1}).  Finding values for
these angles can lead to the determination of the intrinsic beam width
and other geometrical properties of the pulsar emission. For example,
knowing $\alpha$ for a range of pulsars gives us clues about the
origin of their magnetic fields and how the pulsars are evolving
\citep{cb86,b92,m93}, while knowledge of $\beta$ leads to information
on beam properties \citep{b90par,m93}.  In addition, theories about
the structure of pulsar beams make predictions for the orientation of
particular pulsars based on empirical and geometrical equations within
each given theory, so independent methods can be used to verify those
predictions and hopefully help distinguish between the models.  In
this paper, we carefully derive and apply our own technique for
determining pulsar orientation angles, and we compare our method with
others' in order to illuminate the strengths and weaknesses of them
all.

\subsection{The Rotating Vector Model (RVM)}

One way to obtain $\alpha$ and $\beta$ is to examine the position
angle of the linearly polarized emission from the pulsar, and indeed
this is the method that we will use below on our own data.  The
S-shaped curve visible in plots of the polarization position angle
vs. longitude (see, for instance, the middle panel of Fig. 3) is
explained by a model proposed shortly after the discovery of pulsars
\citep{rc69} in which the position angle follows the rotation of the
magnetic field lines at the sub-Earth point on the pulsar.  Called the
Rotating Vector Model (which we denote hereafter as RVM), the model
gives the polarization position angle, $\psi'$, as a function of
pulsar longitude, $\phi$, where one pulsar rotation equals
$360\arcdeg$ of longitude:
\begin{equation}
\tan (\psi' - \psi'_{0}) = \frac{\sin \alpha \sin (\phi - \phi_{0})}
{\sin \zeta \cos \alpha - \cos \zeta \sin \alpha \cos (\phi - \phi_{0})}.    
\end{equation}
\setcounter{footnote}{0}
%\placefigure{fig-1}
(We use $\psi$ for the observed polarization position angle, which
will be shown below to be different from $\psi'$, below.)

The offset angles $\psi'_{0}$ and $\phi_{0}$ (constant for each
pulsar) give the polarization position angle and longitude,
respectively, of the symmetry point and maximum gradient of the
position angle curve (when the pulsar beam is pointed closest to the
observer), and also include arbitrary constant offsets associated with
observing parameters.  The quantity $\alpha$ is the angle between the
positive spin axis (which points in the direction of the angular
velocity vector $\vec{\Omega}$) and the observable magnetic pole,
while $\zeta$ is the angle between the positive spin axis
$\vec{\Omega}$ and the pulsar--observer line of sight.  The sign and
magnitude of $\beta$, the impact parameter of the line of sight with
respect to the magnetic axis, are defined by
\begin{equation}
\beta = \zeta -\alpha.
\end{equation}
In the context of these definitions, $\beta$ is \textit{positive}
whenever $\zeta > \alpha$ so that the line of sight vector is
\textit{farther} from the (positive) spin vector $+\vec{\Omega}$ than
is the observable magnetic axis; and $\beta$ is \textit{negative}
whenever $\zeta < \alpha$ so that the line of sight vector is
\textit{closer} to the (positive) spin vector $+\vec{\Omega}$ than is
the observable magnetic axis.  In our RVM fits that follow, we hold to
these definitions regardless of whether $\alpha$ is greater or less
than $\pi / 2$, so that we remain true to the conventions of Eq. 1.
It is important to note that a positive $\beta$ corresponds to an
``outer'' (i.e., equatorward\footnote{``Equatorward'' indicates that
the line of sight is {\em{opposite}} the spin pole lying nearest to
the observable magnetic pole.}) line of sight only if $\alpha < \pi /
2$; whereas for $\alpha > \pi / 2$ the equatorward line of sight has a
negative $\beta$.  Other investigators use different conventions for
the sign of $\beta$, as we will discuss below.  Furthermore, as
pointed out by Damour \& Taylor (1992)\nocite{dt92} and Arzoumanian et
al. (1996)\nocite{a96}, Eq. 1, used by essentially all researchers who
have fit the RVM model to data, was derived with the convention that
the polarization position angle, $\psi'$, increases \textit{clockwise}
on the sky.  This is contrary to the usual astronomical convention
that \textit{measured} polarization position angle $\psi$ increases
\textit{counterclockwise} on the sky.  Since most previous analyses
have fitted the RVM via Eq. 1 and its clockwise $\psi'$ to data having
the observers' convention of counterclockwise $\psi$ without
correction, we have modified many earlier investigators' results in
order to be consistent with the assumptions of the RVM model.  In what
follows, we refer to this issue as the ``$\psi$ convention problem.''
An RVM fit with the $\psi$ convention problem must be corrected by
transforming the published values, which we refer to as
$\alpha_{orig}$ and $\beta_{orig}$, to values $\alpha$ and $\beta$
that adhere to the clockwise--$\psi'$ RVM rotation convention of
Eq. 1, which we also use in our fitting process:
\begin{equation}
\alpha = 180 \arcdeg - \alpha_{orig}
\end{equation}
\begin{equation}
\beta = - \beta_{orig}
\end{equation}
We will consider this conversion in more depth when we compare our
results to those from other workers.  At that point we will also show
that definitions of some other quantities in some earlier works must
also be modified for consistency with Eq. 1 and its associated
definitions (see Table~\ref{tbl-1} for conversion relationships).

%\placetable{tbl-1}

The $\psi$ convention problem manifests itself in another manner as
well.  One of the most characteristic properties of the RVM is the
steep swing of polarized position angle as the line of sight makes its
closest approach to the magnetic pole.  The slope of this sweep
contains important geometrical information that is used either
explicitly or implicitly by essentially all investigators:
\begin{equation}
\frac{d\psi'}{d\phi}|_{max} =\frac{d\psi'}{d\phi}|_{\phi=\phi_{0}} =
 \frac{\sin \alpha}{\sin \beta}.
\end{equation}
Note that since the observers'--defined $\psi=-\psi'$, a minus sign
must be inserted in the above equation if $\psi'$ is replaced by
$\psi$.  Note also that Eq. 5 fixes the sign of $\beta$ since $\sin
\alpha > 0$.  However, as emphasized in the discussion following
Eq. 2, the sign of $\beta$ does not by itself select inner versus
outer line of sight trajectories. See Table~\ref{tbl-2} for a summary.

%\placetable{tbl-2}

\subsubsection{The Second Magnetic Pole}

The second magnetic pole is relevant in some cases of interpulse
emission. The opposite magnetic pole's colatitude is $\alpha_2 = \pi -
\alpha$, and its impact parameter with respect to the line of sight is
$\beta_2 = \zeta - \alpha_2$.  It is important to note that the RVM
itself does not distinguish one-- from two--pole emission, as the
model provides the position angle of the particular magnetic field
line that happens to be at the sub--Earth point at any instant, which
is a function of magnetic {\em{dipole}} geometry alone.

\subsection{Earlier RVM Fits}

Many researchers have attempted to fit the RVM function to
polarization data (the most comprehensive of which include
\citet{nv82,b91,hx97ab}) in order to determine $\alpha$ and $\beta$.
In comparing those results to our data, however, we have found that
there has been a wide range of definitions of geometrical beam angles
among the different authors. We now discuss their procedures in
detail. In order to compare the different results, we also present for
the first time a dictionary and procedure for converting among the
various investigators' definitions (see Table 1).  The earlier
results, rationalized by the rules of Table 1, are shown in
Table~\ref{tbl-3} along with our results.

\subsubsection{Narayan \& Vivekanand (1982) (NV82)}
\citet{nv82} fitted the RVM model to single--pulse polarization
histograms, thereby eliminating problems caused by emission of
orthogonal polarization modes (see below and \cite{br80}).  They
unweighted those longitudes where the position angle could be
determined in less than 15 percent of the pulses, and uniformly
weighted the rest.  They emphasized the difficulty of distinguishing
outer (i.e., equatorward) and inner line--of--sight trajectories from
RVM fits when the pulsar's emission occupies only a few percent of the
pulse period, as is usually the case.  Despite this difficulty, their
fits did tend to favor one trajectory over the other in most cases.
Their $\alpha$ was measured with respect to the nearest spin pole so
that it is never greater than $\pi/2$ in their full RVM fits, and they
assigned outer (equatorward) lines of sight to positive $\beta$ in all
cases.  These definitions, eminently defensible on physical grounds,
are nonetheless at odds with those of Eq. 1.  Substantial gymnastics
are needed to transform from one to the other.  Uncertainties were
given for only some of the results, and it was not mentioned whether
the strong covariance of $\alpha$ and $\beta$ (which we find in our
own fits, and elaborate on below) is reflected in these uncertainties.

\subsubsection{Blaskiewicz et al (1991) (BCW91)}
These authors fitted the RVM to a subset of the \citet{w99} Arecibo
1418 MHz pulsar polarization data and also searched for manifestations
of special -- relativistic effects.  Publishing while the experiment
was still in its data--acquisition phase, they necessarily used
shorter total integrations than the data analyzed and displayed here.
They chose to exclude data with $L$ below 5 or 10 times its off-pulse
RMS as well as regions appearing to contain orthogonal modes (see
\cite{br80}).  Also, they used \textit{uniform} weighting for the data
that survived to be fitted.  In the end, out of 36 fits on 23 pulsars
at 0.43 and 1.4 GHz, only 17 had uncertainties in $\alpha$ that were
$\lesssim50\arcdeg$. Fractional uncertainties in $\beta$ were often
also very large.  Again, it was not mentioned whether the covariance
of $\alpha$ and $\beta$ is reflected in these errors.  Their solutions
are reported here after correcting for the $\psi$ convention problem.
We will see below that many of our results have smaller uncertainties
than BCW91, presumably because of our larger quantity of data and
consequently improved signal--to--noise ratios, our use of
non--Gaussian statistics and non--uniform weighting in the fitting
procedure.

\subsubsection{von Hoensbroech \& Xilouris (1997a,b) (HX97a,b)}
More recently, \citet{hx97ab} also attempted RVM fits to polarization
data at 1.4, 1.7, and 4.85 GHz. They used the Simplex algorithm to
approach the $\chi^2$ minimum and then zeroed in with the
Levenberg--Marquart fitting algorithm, much as we do below.  Their
quoted uncertainties are also generally rather large, which they
attribute partially to the high correlation of $\alpha$ and $\beta$.
Goodness of fit was assessed with the $\chi^2$ test, with a study of
the departure from Gaussian statistics of the post--fit residuals, and
through evaluation of the symmetry of post--fit residuals around the
best fit.  There is no mention of a threshold noise cutoff nor of
differential weighting of the points.  In Tables 1 and 3, we correct
their results for the $\psi$-- convention problem, and for cases
having their $\alpha_{HX} < 0$.

\subsection{ Previous Empirical / Geometrical (E/G) Solutions for $\alpha$ and 
$\beta$}

Combinations of empirical and geometrical approaches have also been
used to  help determine beam alignments.  In these cases, a full fit to
the polarized position angle curve is {\em{not}} performed.  Instead, 
certain empirical relationships are derived and combined with beam geometry
equations.  The resulting sets of equations are then evaluated as a function
of observed pulsar properties.  By their nature, these techniques do not
yield  estimates of uncertainties.

\subsubsection{Lyne \& Manchester (1988) (LM88)}

Lyne \& Manchester (1988)\nocite{lm88} determined $\alpha$ and $\beta$
for pulsars which they believe exhibit emission from both sides of a
circular cone, having a full longitude width at $10\%$ maximum of
2$\Delta\phi$.  It can be shown geometrically that the emission cone
intrinsic angular radius, $\rho$, is given by
\begin{equation}
\sin^{2}(\frac{\rho}{2}) = \sin^{2}(\frac{\Delta\phi}{2})\sin \alpha \sin(\alpha+\beta) + \sin^{2}(\frac{\beta}{2}).
\end{equation}
LM88 calculated hypothetical perpendicular rotator parameters
$``\beta_{90}$'' and $``\rho_{90}$'' for each such pulsar, where
$\beta_{90} = \beta|_{\alpha=90\arcdeg}$ and $\rho_{90} =
\rho|_{\alpha=90\arcdeg}.$ From Eq. 5, $\sin\beta_{90} =
[\frac{d\psi'}{d\phi}|_{max}]^{-1}$; and Eq. 6 yields
$\sin^{2}(\frac{\rho_{90}}{2}) = \sin^{2}(\frac{\Delta\phi}{2})
\cos\beta_{90} + \sin^{2}(\frac{\beta_{90}}{2})$.  A plot of
$\rho_{90}$ as a function of pulsar period, $P$, reveals a lower limit
to the scatter of data points.  LM88 argue that the lower limit
represents pulsars that truly have $\alpha=90\arcdeg$, in which case
the intrinsic beam radius $\rho=\rho_{90}$, whereas the points
scattered above the limit represent the case $\alpha \neq 90\arcdeg$
so that the true beam width is $\rho< \rho_{90}$. From the lower limit,
they derive the important result:
\begin{equation}
\rho = 6\fdg5 P^{-1/3}.
\end{equation}
Calculating $\rho$ from this relation, and using the other measured
quantities $d\psi'/ d\phi|_{max}$ and 2$\Delta\phi$, it is then
possible to use (the unapproximated) Equations 5 and 6 to calculate
$\alpha$ and $\beta$.

These calculations rest on the assumption that the pulsar beam is
circular (which is backed up by the favorable comparison of their
calculations for the range of the position angle swing (2$\Delta\psi$)
and their measurements of that swing), as well as their idea that
certain pulsars have ``partial cone'' emission for which the above
relationship cannot be used.  Concerning the first assumption, there
has been much controversy over the shape of pulsar beams, with some
investigators advocating emission elongated in the latitude direction
\citep{j80,nv83a}, others finding extension in the longitude direction
\citep{b90par}, \citep{m93}, and yet others agreeing with LM88 that
the beam is essentially circular \citep{b98par}.  It should be noted
that later work by this group \citep{mhq98par,tm98par} suggests a
different functional form for Eq. 7, which if applied to the pulsars
considered here, would yield somewhat different values for $\alpha$
and $\beta$.  Also, it has been argued by \citet{r90} that some of the
types of pulsar emissions under consideration here could be what she
calls ``core'' beams, for which she finds the RVM, and hence Eq. 5,
problematic.

The results from Lyne \& Manchester (1988)\nocite{lm88} were very
helpful to us as initial guesses for our own fits.  However, these
investigators tabulated only $|\beta|$ and confined $\alpha$, by
definition, to $0\arcdeg<\alpha<90\arcdeg$.  We defined the sign of
their $\beta$ from the sign of the position angle sweep ({\it{cf.}}
Eq. 5), but for $\alpha$ we had to choose between $\alpha_{LM}$ and
$\pi - \alpha_{LM}$ solely on the basis of their consistency with our
final results.  Note that this represents the choice between inner and
outer lines of sight trajectories, so it is not trivial (see Section
1.1 and Table~\ref{tbl-2} for further discussion of this issue).

It should be noted that LM88 also provided, in addition to their E/G
analyses, RVM fits to several pulsars having interpulse emission.  We
will discuss these results in the sections on those individual
interpulsars. 

\subsubsection{Rankin (1990, 1993a,b) (R90, R93a,b)}

In a series of papers, Rankin (1983, 1990, 1993a,b) laid out a
comprehensive pulsar classification scheme that distinguishes two
basic types of emission: core pencil beams and hollow cones.  R90
discovered a simple relationship between intrinsic core pulse width
and period by studying pulsars thought to have opposite--pole
interpulses.  Some simple geometry then permits the calculation of
$\alpha$ for any pulsar with a core component.  In this method,
$W_{core}$, the longitude FWHM of an interpulsar with a core
component, was measured and interpolated to 1 GHz.  In a remarkably
good fit, six interpulsars all follow the relation

\begin{equation}
W_{core} = 2\fdg45 \ P^{-1/2}.
\end{equation}

Presumably then for these pulsars, which must have
$\alpha\approx90\arcdeg$ in order that both pulse and interpulse
emission from opposite poles be observable, $W_{core}$ measures the
{\em{intrinsic}} width of the core beam.  All other core pulsars have
larger observed $W_{core}$ for a given period, which is consistent
with the idea of a beam of the same {\em{intrinsic}} width having a
larger longitude extent $W_{core}$ if directed away from the equator
(i.e., $\alpha \not= 90\deg$):

\begin{equation}
W_{core} = 2\fdg45  \ P^{-1/2}/\sin \alpha.
\end{equation}

R90 then took measured values of $W_{core}$ and $P$ to solve for
$\alpha$ via Eq. 9.  Strictly speaking, this procedure yields only
$\sin \alpha$, so that it is not possible to distinguish $\alpha$ from
$\pi - \alpha$.  It should also be noted that this equation assumes
that the impact parameter $|\beta|$ is small, and ignoring $\beta$ is
defended on the grounds that its effect on core components is weak
because the angular intensity may be approximated by a bivariate
Gaussian, whose FWHM is independent of $\beta$.  It is also assumed
that the core emission beam completely fills the open field lines, so
that the core width is directly related to the angular width of these
field lines.  R90 did not attempt to determine $\beta$ from these core
components, stating that ``the polarization--angle behavior of core
components seems to provide no reliable information about the impact
angle $\beta$.''

In a later pair of papers, \citet{r93} did use polarization position
angles to determine $\beta$ for pulsars having {\em{conal}} emission.
The position angle swing relationship, Eq. 5, was used, so both
$d\psi' / d\phi|_{max}$ and $\alpha$ must be determined. The slope was
determined from the observed position angle curve. For pulsars having
both core and conal emission, $\alpha$ was calculated as above, while
other techniques were used in the absence of core emission. R93a,b
attempted to distinguish outer and inner line of sight trajectories by
noting whether the position angle curve did or did not flatten in the
wings ({\em{cf.}} \citet{nv82}), but this was not based on rigorous
fits.  Her sign convention then followed NV82 (positive $\beta$ for
outer line of sight trajectories), although we found that we
occasionally had to flip her published signs for internal consistency
(see Table~\ref{tbl-3}).

Interestingly, R93a,b determined that cones, like cores, also have
intrinsic widths that depend only on period.  Two angular radii are
indicated, suggesting the existence of both an inner and an outer cone
(not to be confused with the inner and outer line of sight
trajectories discussed above).

\subsection{Summary}

The various efforts at RVM fits involve subtle assumptions regarding
choice and weighting of data.  Both of the empirical / geometrical
methods also involve varied assumptions, many of which reach to the
heart of theories of pulsar emission.  Interestingly, although the two
E/G techniques are based on different assumptions and different
definitions, they generally yield similar results for
$\alpha\lesssim40\arcdeg$(\cite{mh93}).

We have fitted our own time-averaged data to the RVM to assess the
efficacy of both empirical--geometrical and RVM approaches to
determining the pulsar and beam orientation.  Work on fitting
single-pulse data is ongoing, and will be addressed in a separate
paper \citep{we2001}.

\section{Observations}

The average pulse data we analyze here were taken at Arecibo
Observatory at 1418 MHz from 1989 through 1993 \citep[(W99)]{w99}.
For these data sets, we attempted to fit the RVM to every pulsar whose
position angle appeared to follow the RVM model.  We were able to find
rigorously convergent fits on only a small fraction of the pulsars in
the database, for reasons that we will discuss below.

\section{Our Fits}

In completing our fits to RVM polarization data, we found that we had
to be especially careful in deciding on statistical vs. uniform
weighting, calculating errors in our polarization data, dealing with
the possibility of orthogonal modes, fitting covariant parameters in
the RVM, and calculating the errors on those covariant parameters.
Each of these concerns is addressed in the subsections below.

\subsection{Statistical Vs. Uniform Weighting}

One of the most important limitations of the RVM fitting technique is
that a large longitude range of polarization position angles is
necessary in order to find trustworthy values for $\alpha$ and $\beta$
\citep{nv82,mh93}.  The narrowness of most pulsar beams unfortunately
renders this goal difficult in most cases.  Thus, many of the best RVM
fits are to pulsars with interpulses, where polarization information
is available at a very large range of longitudes (e.g.,
\citet{lm88,p90}).  This difficulty is aggravated by the fact that
most investigators have \textit{uniformly} weighted all data used in
an RVM fit, eliminating any that falls below some sensitivity
threshold or outside of some longitude range, and therefore reducing
further the amount of polarization information remaining to be
analyzed.  In order to use as much data as possible and therefore
decrease the uncertainty in our fits as much as possible, we
implemented a statistical weighting scheme.  This procedure enables us
to sample a larger range of longitudes and to use $\chi^{2}_\nu$ to
asses the quality of the fit.  However, to use statistical weighting
to its fullest advantage, one must be very careful in calculating
uncertainties for each data point.  It is also important to bear in
mind that {\em{systematic}} errors cannot be accounted for in this
scheme.

\subsection{Error Calculations for Polarization Data}

For the Stokes parameters $I$, $Q$, $U$, and $V$, each associated
standard deviation obeys Gaussian statistics, with
$\sigma_{I}=\sigma_{Q}= \sigma_{U}=\sigma_{V}$. However, polarized
quantities derived from the Stokes parameters such as linearly
polarized power $L$ and position angle $\psi$ are {\em{not}} normally
distributed, and $L$ itself suffers a bias.  In order to correctly use
polarization data, we must take these issues into account.  We show
first how we removed the bias in $L$, and then we discuss the
non--Gaussian errors in $\psi$.  (Errors in $\psi$ and in $\psi'$ are
identical, so for simplicity we will refer only to $\sigma_{\psi}$ in
what follows.)

\subsubsection{Removing the Bias from the Measured Linear Polarization $L_{meas}$}

To compute $L_{meas}$, we used the standard definition:
\begin{equation}
L_{meas} = \sqrt{Q^{2} + U^{2}}.
\end{equation}

The $L_{meas}$ calculated in this way is biased \citep{nc93}, however,
because it is defined as a positive definite quantity and the simple
equation above overestimates the true polarization in the presence of
noise.  This can be seen by noting that if the true values of $Q$ and
$U$ in the above equaton for $L$ were both zero, noise on both of
those values would lead to a {\em{positive}} value for $L_{meas}$.
Thus, the error in the Stokes parameters always increases the value of
$L$, and it is this non-zero bias that we wish to remove from our
value of $L_{meas}$.

\cite{ss85} rigorously compared the merits of various schemes for
debiasing $L$, and showed that several of the estimators lead to
negligibly different outcomes, especially at intermediate
signal--to--noise ratios.  We chose the estimator of \cite{wk74} for
all but the lowest signal strengths, where we set the unbiased linear
polarization to zero following \cite{ss85}.  We first calculated the
off--pulse standard deviation of the total power Stokes parameter $I$,
$\sigma_{I}$.  The measured quantity $L_{meas}/\sigma_{I}$, which can
be considered a kind of ``signal--to--noise,'' then serves as the
correction parameter in the expression for finding the unbiased linear
polarization $L_{true}$:

\[ L_{true} = \left\{ \begin{array}{ll}
             \sqrt{  (\frac{L_{meas}}{\sigma_{I}})^2 - 1} \ \sigma_{I}  & \mbox{if $\frac{L}{\sigma_{I}  } \geq 1.57$ } \\
                 0 & \mbox{otherwise.}  
	        \end{array} 
	  \right. \]

\subsubsection{Error Estimation for Polarized Position Angle $\psi$}

At high signal--to--noise ($S/N$) ratios, the uncertainty in the
position angle is easily calculated:
\begin{equation}
\sigma_{\psi} = \frac{\sigma_{I}}{2 L} = 28\fdg65 \ \frac{\sigma_{I}}{L}
\end{equation}
However, because of the above--discussed bias to the linear
polarization, repeated measurements of the position angle of
polarization at low-- to intermediate--$S/N$ are not normally
distributed, and special formulae must be used to to determine
$\sigma_{\psi}$.  The probability distribution of the position angle
($\psi$) around the true value ($\psi_{true}$) can be calculated
\citep{nc93}:
\begin{equation}
G(\psi;\psi_{true},P_{0}) = \frac{1}{\sqrt{\pi}}\{{\frac{1}{\sqrt{\pi}} + \eta_{0}e^{\eta_{0}^{2}}[1+\mbox{erf}(\eta_{0})]}\}e^{-(\frac{P_{0}^{2}}{2})}
\end{equation}

where $\eta_{0}$ = $\frac{P_{0}}{\sqrt{2}}\cos 2(\psi-\psi_{true})$,
$P_{0}$ = $\frac{L_{true}}{\sigma_{I}}$ (note then that $L$ must be debiased
as shown above before $P_{0}$ can be calculated), and ``erf'' is the 
Gaussian error function.  

Following \citet{nc93}, we define ``$1\sigma_{\psi}$'' via the
numerical integral of the above function (with $\psi_{true}$ = 0) by
adjusting the bounds of integration to include $68.26\%$ of the
distribution, as below:
\begin{equation}
\int_{-1\sigma_{\psi}}^{1\sigma_{\psi}} G(\psi, P_{0}) d\psi = 68.26\%
\end{equation}
To find the error on any given data point, we first built a table of
the results of this numerical integration for $P_{0}$ in the range of
0.0 to 9.99 in steps of 0.01.  Then, to find the ``$1\sigma_{\psi}$''
error bars for a data point having $P_{0} < 10$, we linearly
interpolated on this table. At higher signal strengths, Eq. 11 was
used.

We used the values of $\sigma_{\psi}$ from this procedure as the
uncertainties for each longitude bin in our fits (and thence as the
weighting factors).

\subsection{Orthogonal Modes}

Pulsar radiation is frequently found to be emitted in two orthogonally
linearly polarized modes at the same longitude.  Variations in
strength between the two modes can severely corrupt a classical RVM
curve.  To deal with this possibility, we used the following
procedures: In those rare cases where a $90\arcdeg$ position--angle
discontinuity suggested that emission was switching from predominantly
one mode to predominantly the other, we inserted $90\arcdeg$
discontinuities into our model as well.  In the more common cases of
RVM disruption due to a slow change in the balance of two orthogonally
emitted modes, we unweighted the affected longitudes in our fits.  We
identified these possible orthogonal modes by referring to single
pulse studies when possible \citep{br80,s84}.

\subsection{Fitting Covariant Parameters}

We then fit the RVM model to $\psi(\phi)$ (and $\sigma_{\psi}(\phi)$)
using mainly the Levenberg-Marquardt algorithm \citep{p92} due to its
ability to work efficiently with non-linear functions and to yield
good estimates of the errors on the parameter values it finds.

During the development of the RVM fitting engine, an RVM simulator was
written to ensure that the fitting code was working properly.  In
order to generate realistic simulated data, we first created a classic
RVM position angle curve, and then added the non--Gaussian noise and
biased $L(\phi)$ [see above for discussion of the noise and biasing].
Our noisy, biased simulated $\psi(\phi)$ and $L(\phi)$ were then fed
into our standard fitting program.  Multiple runs of the simulator and
fitting program demonstrated that we could consistently find
$\chi^{2}_{\nu} \approx 1 \pm 0.20$, even with occasional (spurious)
position angle data points far from the main pulse passing the $S/N$
criteria. Comparisons of these advanced simulator runs with other
editions of the simulator (without noise and debias code included)
showed us that even considering the non-Gaussian distribution of
position angle noise and biasing of off-pulse $L$ noise, it should
still be possible to use $\chi^{2}_{\nu}$ as a reasonable
goodness--of--fit indicator.

For initial estimates for the parameters, we used our own best guesses
as well as the results from \citet{nv82,lm88,r90,b91}, and
\citet{r93}.  As outlined in Sections 1.2 and 1.3, however, the
results given by other researchers for $\alpha$ and $\beta$ (or
whatever symbols they used for those angles) were not always
consistent with the RVM equation, and had to be converted to a
consistent definition (see Table~\ref{tbl-1}).

Even with those initial guesses in hand, fitting the data is a very
sensitive and difficult process.  The chief difficulty is that
$\alpha$ and $\beta$ are extremely covariant: without sufficient data
in the ``wings'' of the polarization profile, the fitting engine is
unable to lock into specific values for $\alpha$ and $\beta$ and would
find $\alpha\rightarrow 180\arcdeg$ and $\beta\rightarrow 0\arcdeg$
(while still remaining consistent with the position angle sweep
relationship, Eq. 5).  This problem is also aggravated when $\psi_{0}$
and $\phi_{0}$ are allowed to vary: the fitting engine can sometimes
lock onto values for $\psi_{0}$ and $\phi_{0}$ and will then be unable
to lower $\chi^{2}$ without pulling ($\alpha, \beta$) toward
(180$\arcdeg$, 0$\arcdeg$).  These problems made fitting difficult,
but not impossible, at least in some cases.  To help alleviate these
problems, we added the Powell fitting algorithm \citep{p92} to refine
our initial guesses before passing those results on to the
Levenberg-Marquardt algorithm.  We normally started by using the
initial values from other researcher's results, and manually tweaking
$\phi_{0}$ and $\psi_{0}$ to fit our data.  We then allowed the
fitting engine to fit for $\alpha$, $\beta$, $\phi_{0}$ and $\psi_{0}$
altogether.

At other times, the fit would never truly converge to an answer, but
would instead indefinitely search for a minimum without finding one.
The reason for the difficulty in fitting can be seen in a plot of the
$\chi^{2}$ topography of a particular pulsar, P0301+19, in Fig 2,
which is representative of most. (See also \cite{hx97a}.) There is a
deep chasm in $\chi^{2}$ space representing all $(\alpha, \beta)$
satisfying the observed position angle sweep via Eq. 5.  Consequently
the best--fitting solution $(\alpha_{best}, \beta_{best})$ lies
somewhere in the bottom of the chasm. Unfortunately, the floor of the
entire valley lies at approximately the same ``elevation,'' resulting
in a difficult--to--find minimum and hence a poorly defined best
fit. Note then that it is even difficult to distinguish something as
basic as inner and outer trajectories.  This is a
graphically--oriented description of the fact that $\alpha$ and
$\beta$ are highly covariant as a result of the similarity of all RVM
curves satisfying a particular value of Eq. 5, which is the limiting
case of Eq. 1 over the small longitude range occupied by most pulsars'
emission.

%\placefigure{fig-2}

Another problem is that many pulsars seem to have RVM--style
polarization curves, but upon closer inspection, we discovered that
the curve was of a slightly incorrect shape, and again, the fit would
not converge.  Due to difficulties like these, out of 70 pulsars for
which we had average pulse position angle data, only ten could be fit
reliably. The reliability of the fit was assessed by using other
researchers' results as initial parameter estimates for our
algorithm. In each of the ten successful cases, all of our fits
converged to the same final results, which often were {\em not} the
same as the other published solutions.  The results for our fits are
listed in Table~\ref{tbl-3}, along with values from other researchers
for those same pulsars, which have all been converted to consistent
definitions for $\alpha$ and $\beta$ via Table~\ref{tbl-1}.

\subsection{Calculating Errors on the Covariant Parameters}

With the output from the fitting engine, we were able to calculate
errors based on the full length of the region lying at
$\Delta\chi^{2}_{\nu}=1$ above the minimum, which is imperative
because $\alpha$ and $\beta$ are highly covariant
\citep{hx97a,p92}. The estimated uncertainties reported in
Table~\ref{tbl-3} reflect the covariant relationship between
parameters.

\subsection{Results}

We describe here  the results of our fits and compare them with other 
investigators' efforts.  The fits are  shown in 
Figs. 3--17, and listed and compared with others' in Table 3.

\subsubsection{PSR B0301+19 (Fig. 3)}

The two other RVM fits at frequencies similar to or below ours, NV82
and BCW91, obtain substantially different results from our fit.  All
three yield similar values of the ratio $\sin\alpha / \sin\beta$, as
expected from Eq. 5.  We find that the other two fits share similar
$\beta$ and appear to have $\alpha$ related by $\pi -
\alpha_{NV}\approx\alpha_{BCW}$ although we believe that we have
reduced all results to common definitions. Note that this
transformation would represent a switch from an inner to an outer line
of sight trajectory. The disparities highlight the great difficulty in
determining a unique $\alpha$ and $\beta$ from an RVM fit for these
two highly covariant parameters.  The HX97a,b fit is at a much higher
frequency and not surprisingly has quite different parameters.

The high quality of our position angle data, representing $0.3 d$ of
integration at Arecibo Observatory, has enabled us to fit farther into
the pulse profile wings than could the earlier RVM investigators.
Apparently this extra longitude range permitted a more robust
solution. It is quite interesting to note that our results are much
closer to the two E/G investigations, which also are similar to each
other.

%\placetable{tbl-3}

\subsubsection{PSR B0525+21 (Fig. 4)}

The results in Table 3 appear rather scattered at first glance.
However, We agree with BCQ to within a few $\sigma$, and the HX97a,b
1.41 GHz result would also agree if one took the complement of
$\alpha_{HX}$ (switching from an inner to an outer line of sight
trajectory). However, the two E/G results and the NV82, 0.43 GHz RVM
fit place $\alpha$ far away, at or above $\sim155\arcdeg.$ We favor
our result, which is based on $1.9 h$ of integration at Arecibo, again
permitting a fit farther into the profile wings.

\subsubsection{PSR B0656+14 (Fig. 5)}

The low--level polarized emission between $(18-30)\arcdeg$ longitude
provides part of the critical second half of the classic S--shaped RVM
curve.  It is this extra longitude coverage, resulting from a deep
$1.8 h$ integration, that enables the fit to converge to a unique
value.  Note that our fit moves $\phi_{0}$, the longitude of symmetry,
to the end of the main pulse component, indicating that the main
component and the trailing low--level emission constitute the opposite
sides of a cone.  There are no other RVM results on this pulsar.  Our
rather large error bars encompass the E/G results.

\subsubsection{PSR B0823+26 (Figs. 6,7)}

Both R93 and W99 have labelled the main component a core, based
primarily on its polarization properties. (R93 called the whole
profile $T$, while W99 labelled it $S_{t}$ based also on newer
high--frequency observations.)  The pulsar presents a main pulse plus
a postcursor at $\phi\sim40\arcdeg$ and a very weak interpulse near
$\phi\sim180\arcdeg$.  An apparent orthogonal mode switch between
$\phi = (-180$ and $-7)\arcdeg$ and $\phi = (+140$ and $+190)\arcdeg$
was included in the fit by adding $90\arcdeg$ to the position angle
over that longitude range, while a region of competing orthogonal
modes was unweighted in the $\phi = (-7$ to $-1)\arcdeg$ range.  We
present two sets of results -- the preferred one to the full main /
postcursor / interpulse component range, and another one to the first
two components alone (see Table~\ref{tbl-4}).  The two results differ
by several times their formal errors, a point to which we shall return
below with pulsar B0950+08.  Note also that our two results place
$\alpha$ in (barely) opposite hemispheres.  This seems rather
unimportant until one realizes that the negative $\beta$ of both fits
then implies an inner line of sight trajectory in one fit, but an
outer trajectory in the other (see Section 1.1 and Table 2 for further
discussion). However, the distinction between inner and outer is less
important for equatorial trajectories such as these.

Our results comport with the other three RVM fits on this pulsar at
similar frequencies: NV82, BCW91, and HX97a,b, although error bars on
their results are generally quite large or not given.  (A 0.43 GHz
BCW91 fit with small error bars is also in quite good agreement with
our 1.4 GHz results.)  Both of the empirical/geometrical main--pulse
results lie close to ours.  LM88 also derive E/G results for the
interpulse, as shown in Table 3. In addition, LM88 crudely fitted the
RVM to a combination of 0.43 and 1.4 GHz position angle data for all
three pulse components, with the result that $\alpha=100 \arcdeg$,
$\beta=-5.2\arcdeg$ (no uncertainties given). The fourth panel of
Figs. 6 and 7, which shows the postion angle calculated from many
workers' results atop our data, demonstrates that our fit yields a
better representation of our data than does any other published
result. Our careful treatment of uncertainties and biases in the
extensive low--level emission also favors our result.

It is clear from our fit that the interpulse represents emission from
the opposite pole having $\alpha_{2}= 81\fdg1$, so
$\beta_{2}=\zeta-\alpha_{2}=14\fdg8$. This large impact parameter
helps to explain why the interpulse is so much weaker than the main
pulse.  The postcursor is then emitted from the vicinity of the
primary magnetic pole but at higher impact parameter than the main
pulse.

%\placetable{tbl-4}

\subsubsection{PSR B0950+08 (Figs. 8, 9, 10)}

In our preferred fit (Fig. 8), we use data from all longitudes, with
the exception of the ranges $\phi = (-10$ to $+15)\arcdeg$.  We also
account for the orthogonal mode emission between $\phi = (25$ to
$175)\arcdeg$.  The resulting fit appears quite robust across
virtually the whole fitted longitude range.

Our result is significantly at odds with all the others (including
both RVM and E/G determinations), which are rougly consistent with
each other. While our solution suggests a two--pole orthogonal rotator
like B0823+26, the other fits (including an LM88 RVM fit to data
stitched together from several sources yielding
$(\alpha,\beta)=(170,4.6)\arcdeg$) place $\alpha\approx170\arcdeg$
with both pulse and interpulse emission from a single pole.
[\citet{nv83b} and \citet{g83} elaborated a one--pole model for this
pulsar.]

Our result indicates that the second pole has $\alpha_{2} =74\fdg6$
and $\beta_{2}=52\fdg9$.  \citet{hc81} showed that the main
pulse--interpulse longitude separation remains constant between 40 and
4850 MHz, as expected of a two--pole emitter.  However, the emission
{\it{between}} the pulse and interpulse components is then
particularly puzzling, as it is emitted even farther from the two
magnetic poles than are the pulse and interpulse themselves.

Our two other solutions select data over narrower longitude ranges to
test the consistency of our fits.  (see Table~\ref{tbl-4} and Figs. 9
and 10).  First, we unweight data in the vicinity of the main pulse,
leaving what appears to be a fine fit across the interpulse and
elsewhere (Fig. 9).  The angles $\alpha$, $\beta$ and $\phi_0$ change
significantly.  The second additional fit unweights all data near the
interpulse.  Note that this ``main pulse'' fit looks quite reasonable
throughout the fitted range (see Fig. 10), yet $\alpha$ and $\beta$
move from our preferred fit by several times the formal errors.

While both of these latter fits appear reasonable over their fitted
ranges, they both fail elsewhere.  Only the first, full longitude
range fit, conforms well to observed position angles at all longitude
ranges.  We believe that our high--quality data, coupled with our
careful treatment of measurement uncertainties in the fits, provide a
more accurate result.  For comparison, we superpose some of the other
workers' results, along with ours, onto our data in the bottom panel
of each of Figs. 8, 9, and 10.  Note that only our full fit
(Fig. 8) matches well the overall slope of position angle with
longitude over the longitude range $\sim-160$ to $-20\arcdeg$. (All
results including ours have some trouble matching the left edge of the
interpulse, presumably as a result of orthogonal mode competition.)

We are confident that our full range fit is the best one. However, our
own three somewhat inconsistent solutions help to illustrate the
pitfalls of RVM fitting.  Since most other pulsars' emission is
detected over a much narrower longitude range, the resulting solutions
could be expected to be no more representative of the actual situation
at the pulsar than are the two restricted range fits discussed here.
Presumably orthogonal mode mixing and/or other systematic effects
cause these inconsistencies.  Clearly the formal uncertainties
generally underestimate the errors, as was also the case for pulsar
B0823+26.

Similarly, the fact that other RVM investigators converged on a much
different solution than we did (indicating a one--pole emitter rather
than our two--pole orthogonal rotator) shows how sensitive the fitting
process is to low--level emission which must be properly weighted.

\subsubsection{PSR B1541+09 (Figs. 11, 12)}

This pulsar has an unusually wide profile, which extends over
$100\arcdeg$ in longitude.  Unfortunately a large region, from
$\sim-10$ to $\sim+35\arcdeg$ longitude, must be excluded in order to
achieve convergence.  Note that within the context of the RVM model,
the only way to explain the reversal of sign of $d\psi/d\phi$ within
the profile is to posit an outer line of sight trajectory, with the
extremum in $\psi$ occuring at a longitude that is some distance from
$\phi_{0}$ \citep{nv82}. Indeed our fit, after excluding the above
noted range, conforms to these conditions. However, the {\em full}
range of observed position angles cannot be fitted to any possible
set of parameters in this scenario.  More likely, emission in the
excluded range represents the combination of two RVM orthogonal modes
with relative amplitude varying across the excluded zone. \citet{br80}
show orthogonal mode competition is important at lower longitudes than
these at 430 MHz; unfortunately there are apparently no 1.4 GHz
orthogonal mode analyses.

There are no other published RVM fits, but the two E/G results (which 
are close to one another) are significantly at odds with ours.  However, 
the position angles derived from both of their
results bear no resemblance to the observed position angles, which
leads us to doubt their efficacy, at least at our frequency of 1.4 GHz
(see the bottom panels of Figs. 11 and 12).

Proceeding under the assumption that some or all of the data at
longitudes below $\sim35\arcdeg$ must be unweighted, we were able to
obtain two convergent fits.  In our preferred fit (see Fig. 11), the
longitude range $(-8,+38)\arcdeg$ is unweighted, but lower longitudes
fit quite well onto the RVM. In the second fit, all longitudes below
$30\arcdeg$ are unweighted.  The two fits yield rather different
results for $\alpha, \beta,$ and $\phi_0$.

In our fits, the inflection point at $\sim 52$ or $\sim62\arcdeg$
longitude represents the closest approach of the magnetic axis and the
line of sight. It is interesting that the {\em very} strong circular
polarization (among the strongest observed in any pulsar) occurs at
longitudes similar to our excluded range, suggesting an unusual
emission component much like a core except that it is not at the
symmetry point, as expected.  W99 reviewed all available
multifrequency profile measurements and concluded that this pulsar is
a classical triple ($T$), also identifying the highly circularly
polarized component with the core.  The current fit differs from their
result in that it finds the symmetry center {\em after} the highly
circularly polarized component, rather than in it.  The leading
component is then even farther from the symmetry center, rather than
lying symmetrically opposite the trailing component.

\subsubsection{PSR B1839+09 (Fig. 13)}

Our result with $\alpha$ nearly $90\arcdeg$ and $\beta$ near
$2\arcdeg$ agrees very well with the E/G analyses of R93 and LM88, in
our fit encompassing $80\arcdeg$ of longitude.  The only other RVM
fit, from BCW91, did not achieve statistically significant results.
Note that we unweighted the orthogonal mode competition starting at
longitude $4\arcdeg$.

\subsubsection{PSR B1915+13 (Fig. 14)}

For this pulsar, we fit to all data within $60\arcdeg$ of longitude
centered around the main pulse. The classical RVM sweep appears to fit
the data well.  Comparing our results to others, the BCW91 error bars
on $\alpha$ and $\beta$ barely overlap with ours. As our fit
is based on significantly more data, it is probably more reliable.
The agreement with HX97a,b at 4.85 GHz is poorer, which is not
surprising given the large frequency difference.  The E/G results of
R93a,b are the closest solutions to ours.

\subsubsection{PSR B1916+14 (Fig. 15)}

We fit the central $13\arcdeg$ of longitude.  The only other RVM fit,
from BCW91, has very large uncertainties which encompass our fit.  Of
the two E/G results, LM88 find $\alpha$ and $\beta$ near ours but the
RVM curve derived from their parameters deviates significantly from
our data, while R93a,b exhibits the opposite situation: her $\alpha$
and $\beta$ are far from ours but the RVM curve derived from her
parameters matches our data well.  The latter situation results from
the strong covariance of the two parameters.

\subsubsection{PSR B1929+10 (Figs. 16, 17)}

For our preferred fit (Fig. 16), we fit all $360\arcdeg$ of
longitude, except that we unweight the main pulse region from
$-17\arcdeg$ to $+10\arcdeg$. The excluded region could not be fitted
satisfactorily with the RVM curve, especially when trying to fit the
interpulse and other off--pulse emission simultaneously with the main
pulse. The single--pulse polarimetry of \citet{rr97} [RR97] shows that
indeed a competition between two orthogonal modes leads to position
angle distortions in this region.  Because the pulsar is so strong at
these longitudes, our weighting scheme would otherwise cause this
region to dominate the fit.  We also placed a $90\arcdeg$ orthogonal
mode offset from longitude $-60\arcdeg$ to $-15\arcdeg$, which is also
evident in the single--pulse displays of RR97. Our second fit
eliminates the main pulse entirely, (see Table~\ref{tbl-4} and
Fig. 17)), and yields results roughly similar to the first.

Our $\alpha$ of approximately $35\arcdeg$ and $\beta \sim 26\arcdeg$
seems to be another shot at what seems a scatter of results for this
pulsar. Other results not listed in Table~\ref{tbl-3} include the
following RVM fits: LM88, using the 430 MHz data of \citet{rb81},
found $(\alpha,\beta) = (15,7.5)\arcdeg$, \citet{p90} [P90] measured
$(\alpha,\beta) = (35\pm4,21\pm3)\arcdeg$ at 1665 MHz and
$(\alpha,\beta) = (30\pm2,20\pm2)\arcdeg$ at 430 MHz, and RR97 found
$(\alpha,\beta) = (31,20)\arcdeg$ at 430 MHz. Note that most of the
RVM fits, including ours, yield roughly similar results, which
indicate that the pulse and interpulse are emitted from nearly
opposite sides of a wide, hollow cone near the $+\vec\Omega$ spin
axis. While one can find the colatitude and impact parameter of the
other magnetic pole as $\alpha_2 = 144\fdg0$ and $\beta_2=-82\fdg5$,
these quantities are not important in a one--pole emission model. The
significantly nonzero value of $\phi_0$ indicates that the brightest
emission does {\em{not}} occur at closest approach of the line of
sight and the magnetic axis.  Aside from his choice of emission
centered on the $-\vec\Omega$ spin axis, we are in agreement with the
rough sketch of the emission geometry shown in P90.

R93, using E/G arguments, also found $(\alpha,\beta) =
(18,11.6)\arcdeg$ if B1929+10 is interpreted as a $cT$--type pulsar,
but suggested that it might be an orthogonal rotator if it is of the
$T$ class.  While the RR97 RVM fit discussed above also suggests a
moderately tipped dipole, the authors argue that RVM fits lead to
spurious results for this pulsar, and they favor an orthogonal rotator
$(\alpha,\beta) \sim (90,0)\arcdeg$, dual pole emission model, again
on the basis of E/G considerations.  We still prefer our RVM results,
as we believe it is unlikely that such a beautiful fit obtained over
such a wide longitude range could be so seriously corrupted by
systematic effects as to render it invalid.

\section{Conclusions }

We have fitted pulsar polarization data to the rotating vector model
(RVM) in order to find geometrical parameters.  We were careful to
account properly for the bias in $L$ at low $S/N$ and for
non--Gaussian statistics of position angle data, so that the low $S/N$
data could be appropriately weighted in the fits. This enabled us to
correctly include low--level emission arising from weak pulse
components as well as the wings of main components in the fits.  We
are not aware of any other RVM fits that include such considerations.

We succeeded in finding convergent RVM fits on ten of the pulsars
observed by \citet{w99}. In all such cases, we always converged to our
same solution regardless of our choice of initial parameters gleaned
from previously published results.  Yet we were surprised that our
fitting procedure converged for relatively few of the many pulsars for
which we had high--quality data.  Presumably this was due to
competition between orthogonal emission modes or other systematic
effects.  Because of the high covariance between the two important
fitted parameters $\alpha$ and $\beta$, even slight deviations of
position angle data from the nominal RVM model could prevent
convergence to a unique solution.  Wide longitude ranges of emission
were generally required in order to achieve unique, convergent fits.
As a result, pulsars with interpulses and/or other off pulse emission
are favored targets for RVM studies.  Nevertheless, we show that in
two such cases (pulsars B0823+26 and B0950+08), our fits to different
longitude ranges yield results differing by several times their formal
errors.  Consequently it is clear that systematic effects are present
at a significant level.

We created transformation equations to convert other published RVM and
empirical--geometrical (E/G) results to a common definition, in order
to compare them with our results and to facilitate future comparisons.
For some pulsars, all workers' results are similar while for others
there is a wide range of solutions.  No clear pattern emerges from
these comparisons.

We are currently in the process of separating orthogonal modes in
single pulse polarization data \citep{we2001}, in order to to
determine if RVM fits to separated emission modes yield results
different than the average pulse results presented here.

\acknowledgements{This work was supported by NSF grant AST--9530710.
JEE additionally acknowledges the support of NASA grant NAG5--9063,
and the hospitality of the Smithsonian Astrophysical Observatory,
where parts of this work were completed.  Arecibo Observatory is
operated by Cornell University under cooperative agreement with the
NSF.}

\newpage
\clearpage
%fig 1 follows:

%fig 2 follows:
\clearpage
\begin{figure*}
\centerline{\includegraphics[width=3.5in, height=3.5in]{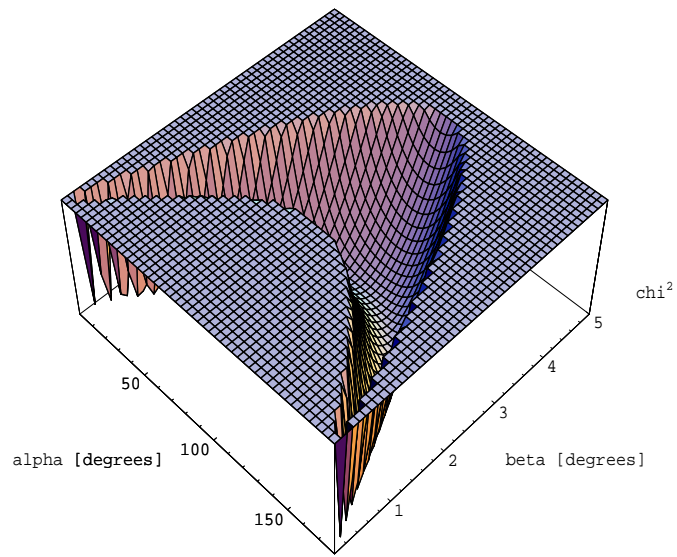}}
\caption{$\chi^{2}$ surface as a function of
$\alpha$ and $\beta$ for pulsar B0301+19  Note that there is a long, deep,
and virtually flat--bottomed 
chasm defined by the observed value of position angle sweep rate 
$d\psi/d\phi$, via Eq. 5.
\label{fig-2}}
\end{figure*}

%fig 3 follows:
\clearpage
\begin{figure*}
\epsscale{0.6}
\plotone{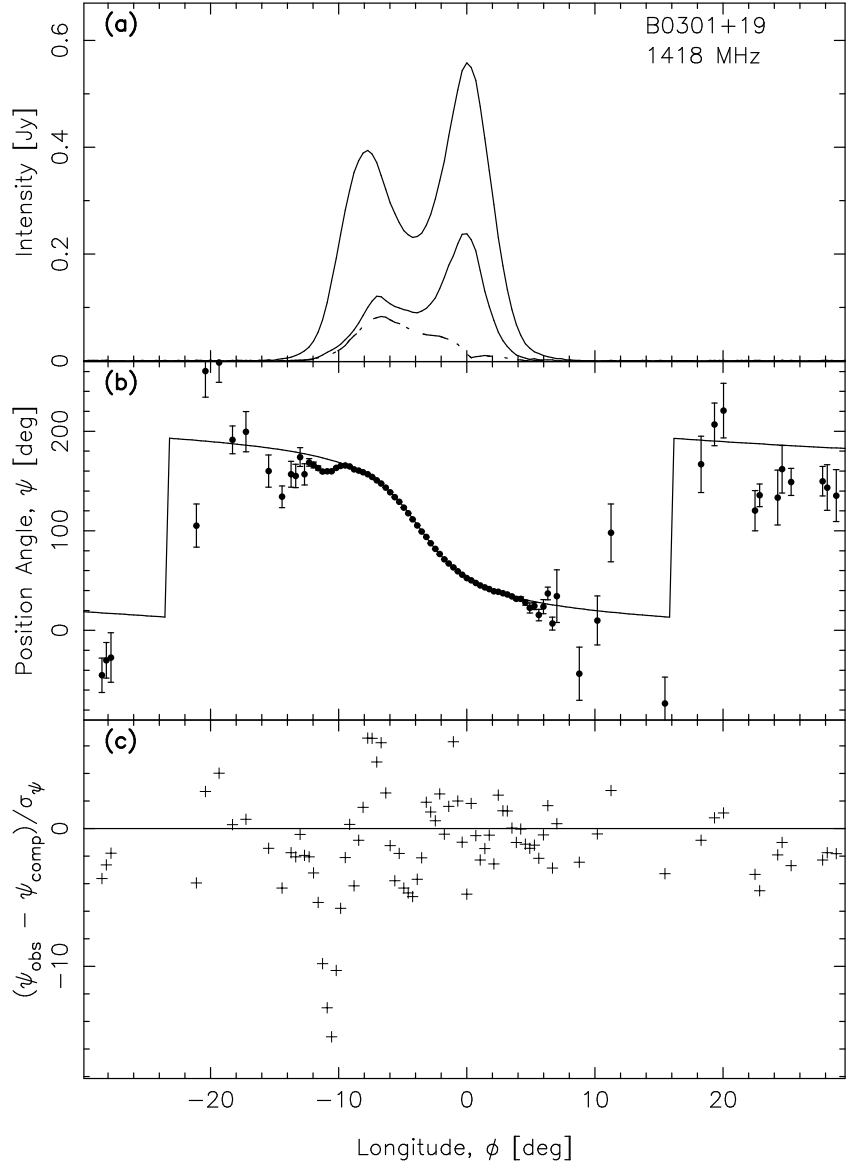}
\caption{Pulsar B0301+19:  (a): Total ($I$), linearly
polarized ($L$), and circularly polarized ($V$, dashed--dotted), flux
densities;  (b): Observed and fitted position angle of
linear polarization $\psi$ with error bars representing measurement
uncertainties; and (c): Position angle residuals,
normalized by $\sigma_{\psi}$.  The fit was allowed to extend over the
full displayed longitude range.
\label{fig-3}}
\end{figure*}

%fig 4 follows:
\clearpage
\begin{figure*}
\epsscale{0.6}
\plotone{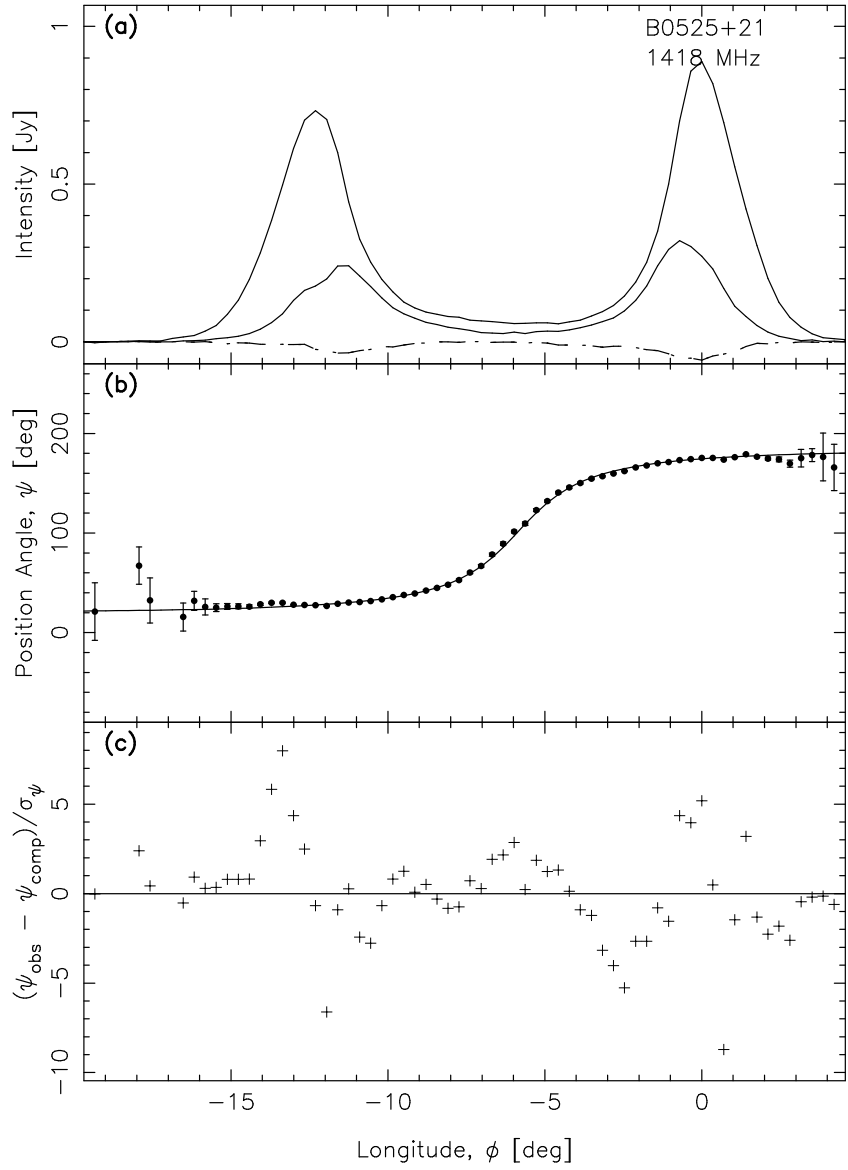}
\caption{Pulsar B0525+21: See Fig. 3 caption for details.
\label{fig-4}}
\end{figure*}

%fig 5 follows:
\clearpage
\begin{figure*}
\epsscale{0.6}
\plotone{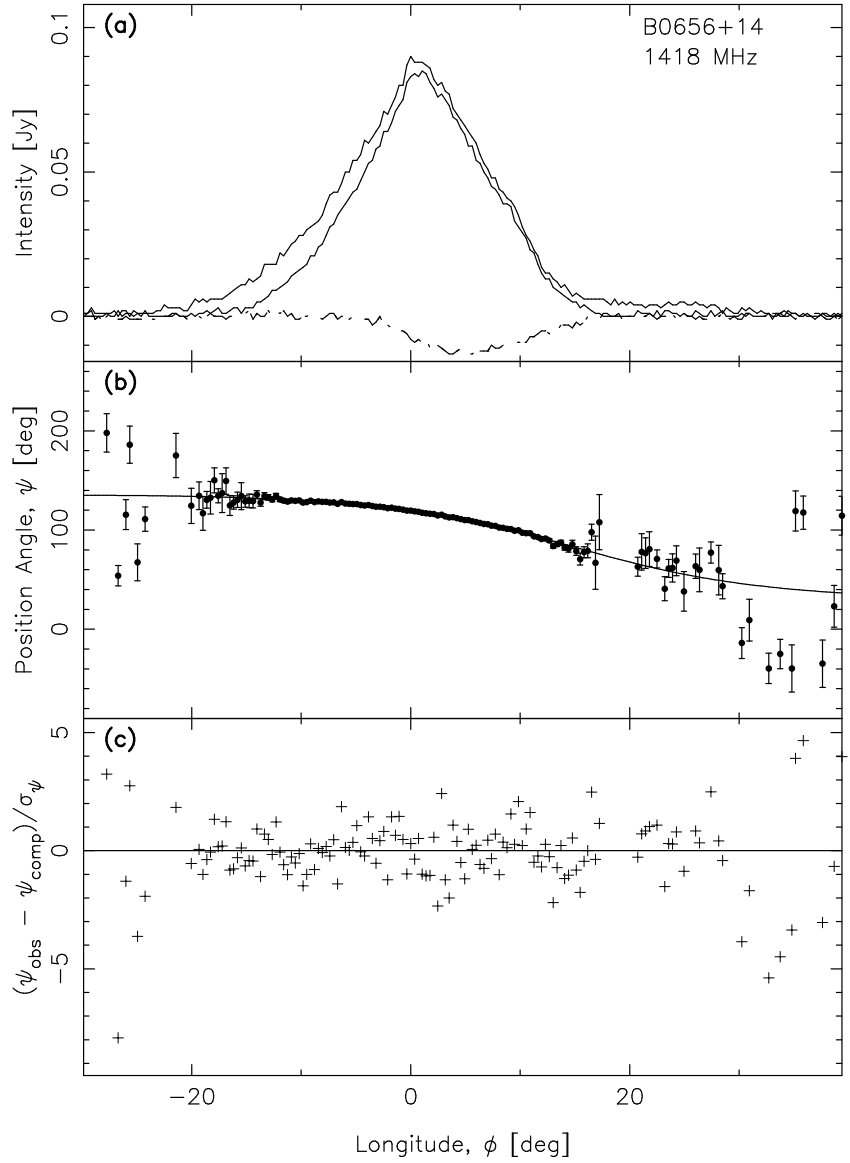}
\caption{Pulsar B0656+14 See Fig. 3 caption for details.
\label{fig-5}}
\end{figure*}

%fig 6 follows:
\clearpage
\begin{figure*}
\epsscale{0.5}
\plotone{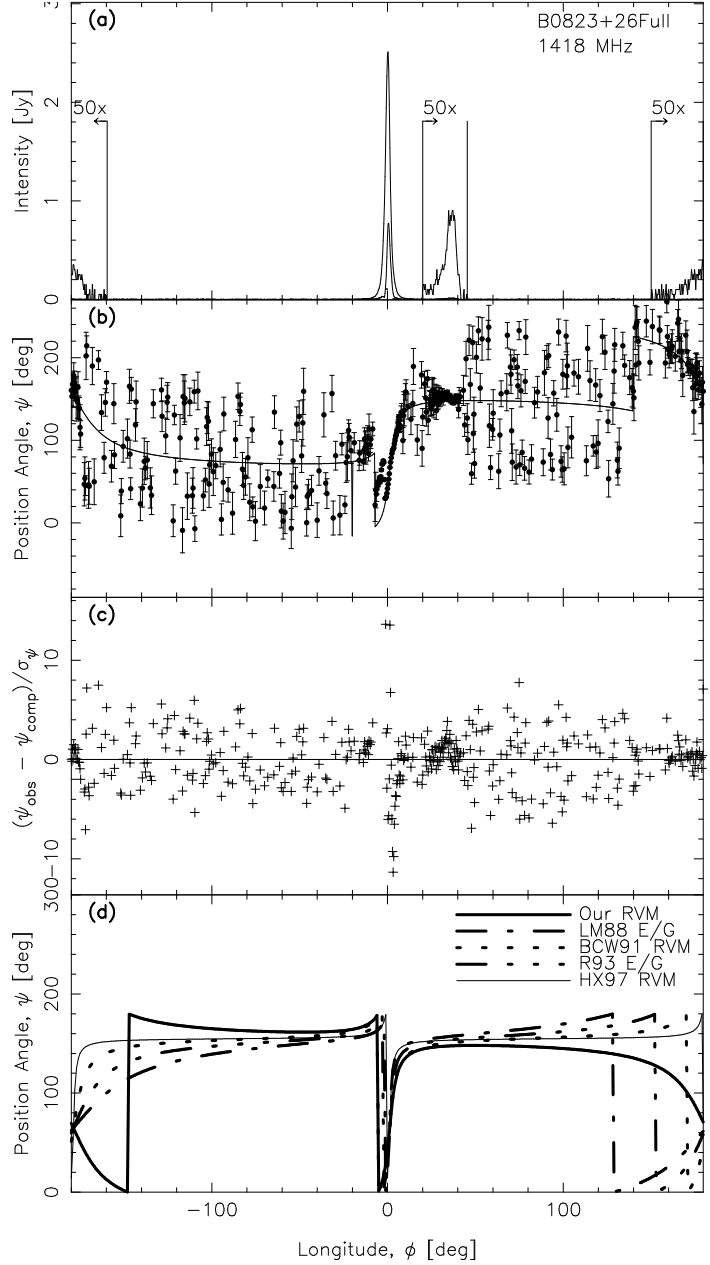}
\caption{Pulsar B0823+26.  Our preferred full longitude fit.  For
an explanation of (a) -- (c), see Fig. 3 caption. Note that a $90\arcdeg$
orthogonal mode switch is inserted in the longitude range
$(-180,-7)\arcdeg$ and $(+140,+180)\arcdeg$, and that longitudes
between $(-7,-1)\arcdeg$ are unweighted in the fit.  Panel (d) shows
the RVM curves from our and other investigators' work.  Orthogonal
mode jumps included in our fit and shown in (b) and implicitly in
residuals (c) are {\em not} included in (d), in order to simplify
comparison with other authors' work.  Also, since other investigators
generally did not provide their fitted position angle and longitude
offsets $\psi_{0}$ and $\phi_{0}$, we use our values of these two
parameters in calculations of their RVM curves.  At
the resolution of panel (d), the NV82 RVM is indistiguishable from
ours, so we do not plot it. 
\label{fig-6}}
\end{figure*}

%fig 7 follows:
\clearpage
\begin{figure*}
\epsscale{0.5}
\plotone{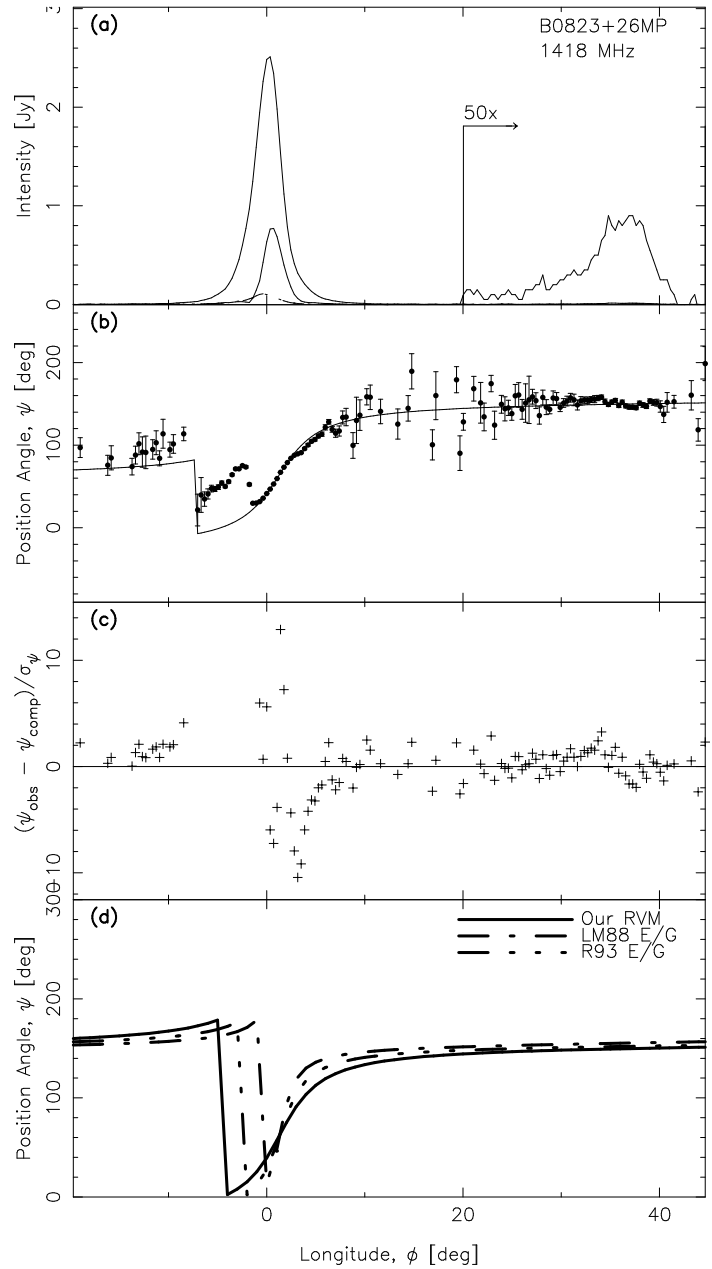}
\caption{Pulsar B0823+26.  Main pulse through postcursor fit.  For
general explanation of this figure, see Fig. 3 and Fig. 6 captions.
Note that a $90\arcdeg$ orthogonal mode switch is inserted in the
longitude range $(-20,-7)\arcdeg$ and that longitudes between
$(-7,-1)\arcdeg$ are unweighted in the fits shown in (b) and (c).  At
the resolution of panel (d), both the BCW91 RVM and the NV82 RVM are
indistiguishable from ours, so we do not plot them.  Also, the HX97 RVM
curve overlays the LM88 E/G curve, so we do not plot the HX97 RVM
fit.  
\label{fig-7}}
\end{figure*}

%fig 8 follows:
\clearpage
\begin{figure*}
\epsscale{0.5}
\plotone{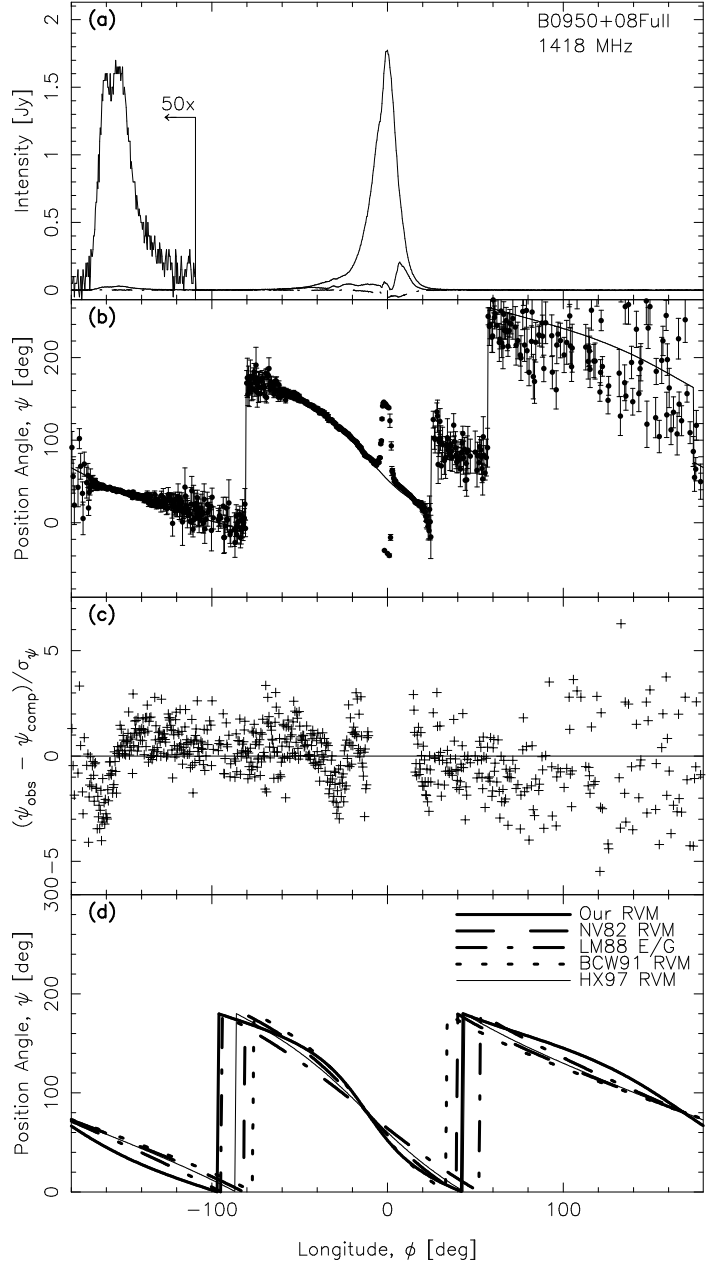}
\caption{Pulsar B0950+08. Our preferred full longitude fit.  For
general explanation of this figure, see Fig. 3 and Fig. 6
captions. Note that a $90\arcdeg$ orthogonal mode switch is inserted
in the longitude range $(25,175)\arcdeg$ and that longitudes between
$(-10,+15)\arcdeg$ are unweighted in the fits shown in (b) and (c).
At the resolution of panel (d), the R93 E/G curve is indistiguishable
from the LM88 E/G curve, so we do not plot the R93 E/G curve.
\label{fig-8}}
\end{figure*}

%fig 9 follows:
\clearpage
\begin{figure*}
\epsscale{0.5}
\plotone{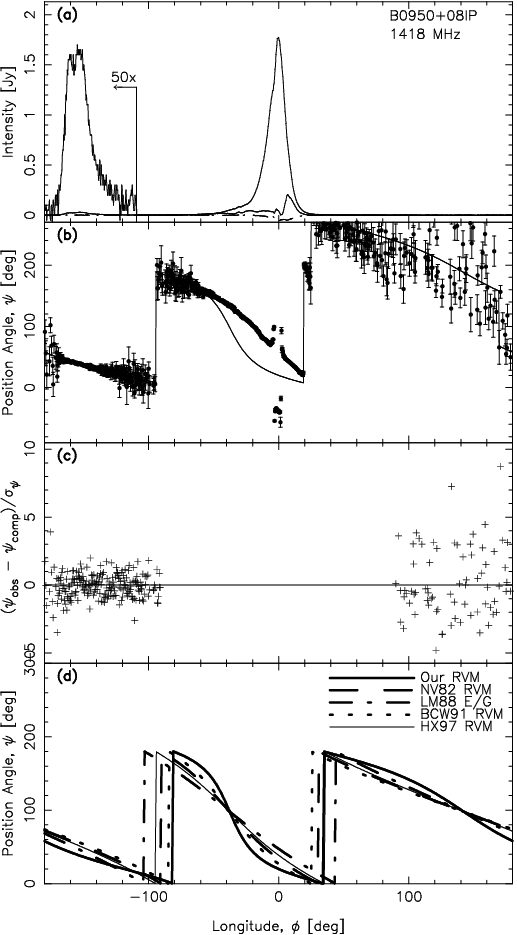}
\caption{Pulsar B0950+08. The interpulse is weighted in this fit, with
the main pulse ($\phi = (-90,90)\arcdeg$) unweighted. For general
explanation of this figure, see Fig. 3 and Fig. 6 captions.  Note that
a $90\arcdeg$ orthogonal mode switch is inserted in the longitude
range $(25,170)\arcdeg$ and that longitudes between $(-90,90)\arcdeg$
are unweighted in the fits shown in (b) and (c).  We omit the same
curves as in Fig. 8. 
\label{fig-9}}
\end{figure*}

%fig 10 follows:
\clearpage
\begin{figure*}
\epsscale{0.5}
\plotone{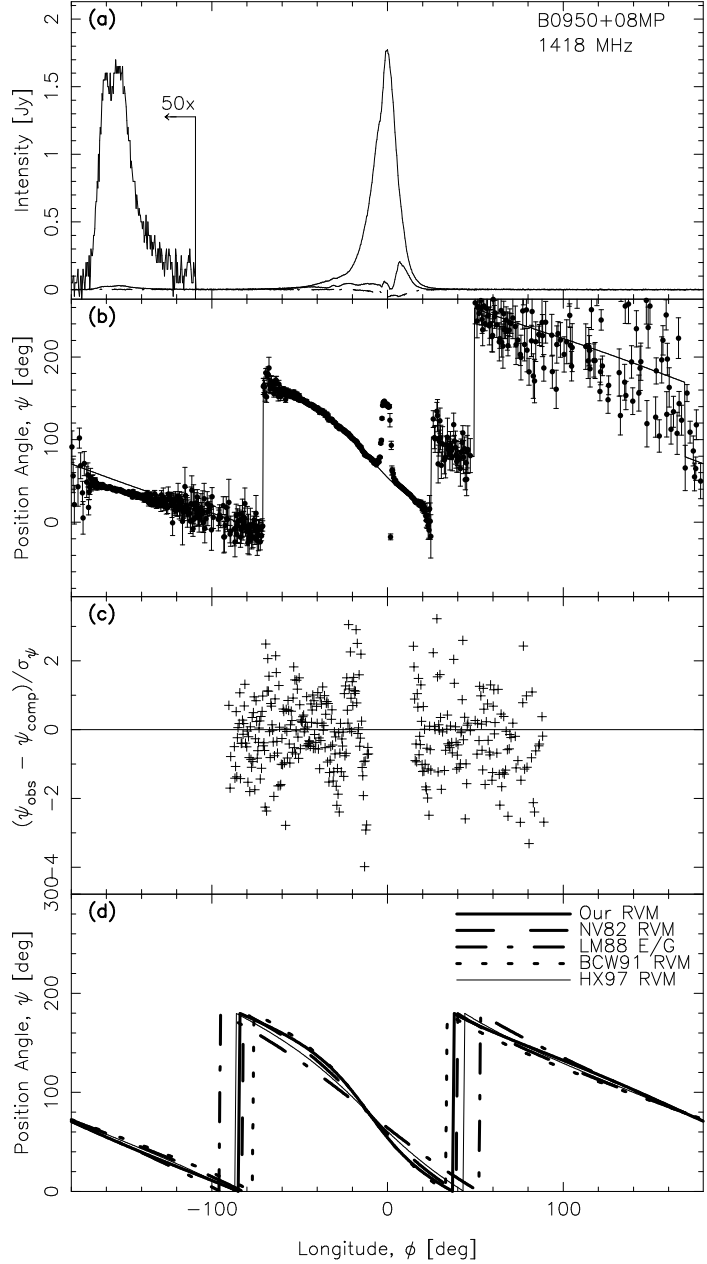}
\caption{Pulsar B0950+08. Main pulse weighted, with the interpulse
unweighted. For general explanation of this figure, see Fig. 3 and
Fig. 6 captions.  Note that a $90\arcdeg$ orthogonal mode switch is
inserted in the longitude range $(25,170)\arcdeg$ and that longitudes
between $(-180,-90)\arcdeg$, $(-10,15)\arcdeg$, and $(90,180)\arcdeg$
are unweighted in the fit shown in (b) and (c).  We omit the same
curves as in Fig. 8. 
\label{fig-10}}
\end{figure*}

%fig 11 follows:
\clearpage
\begin{figure*}
\epsscale{0.5}
\plotone{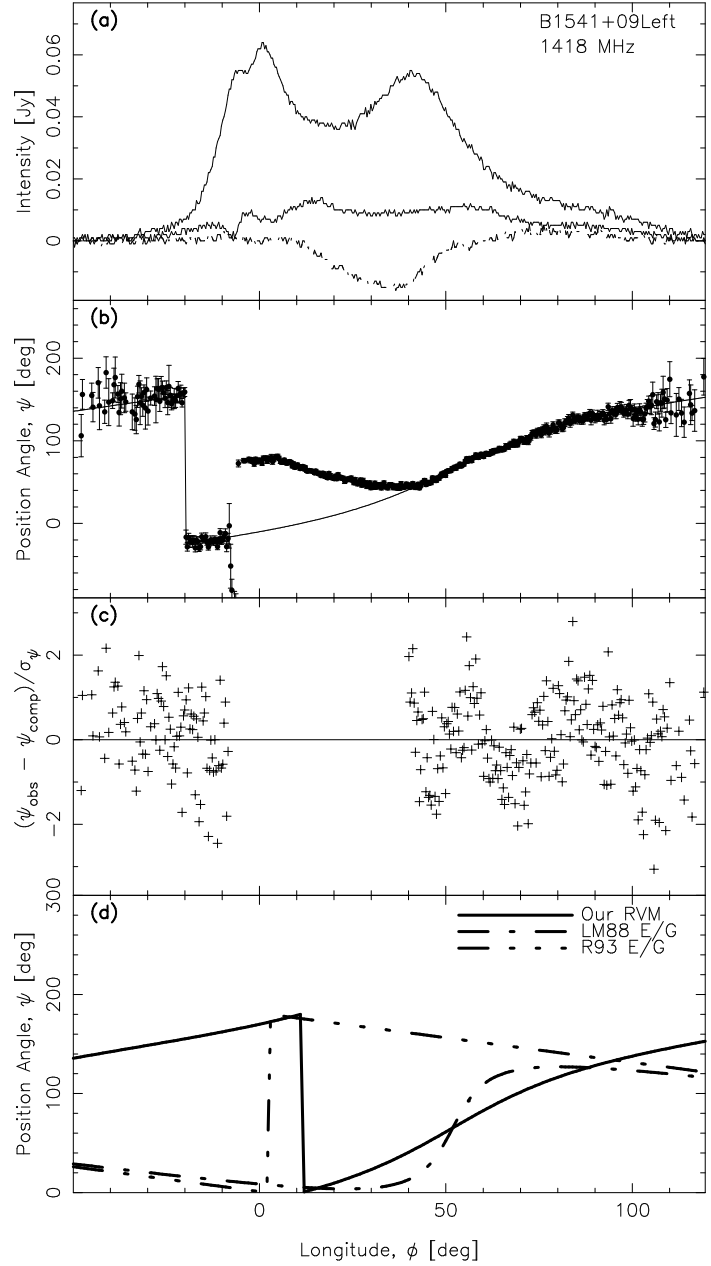}
\caption{Pulsar B1541+09. Our preferred wide longitude fit. For
general explanation of this figure, see Fig. 3 and Fig. 6 captions.
Longitudes between $(-8,+38)\arcdeg$ are unweighted in the fit shown
in (b) and (c). The R93 curve in panel (d), which appears to have the
wrong slope, corresponds to $(\alpha,\beta)=(175,0)$, which represents
a line of sight traverse right across the magnetic pole.  Even a
slight nonzero $\beta$ would add a region of inverted slope near
$\phi=\phi_0$, which would better match our data.
\label{fig-11}}
\end{figure*}

%fig 12 follows:
\clearpage
\begin{figure*}
\epsscale{0.5}
\plotone{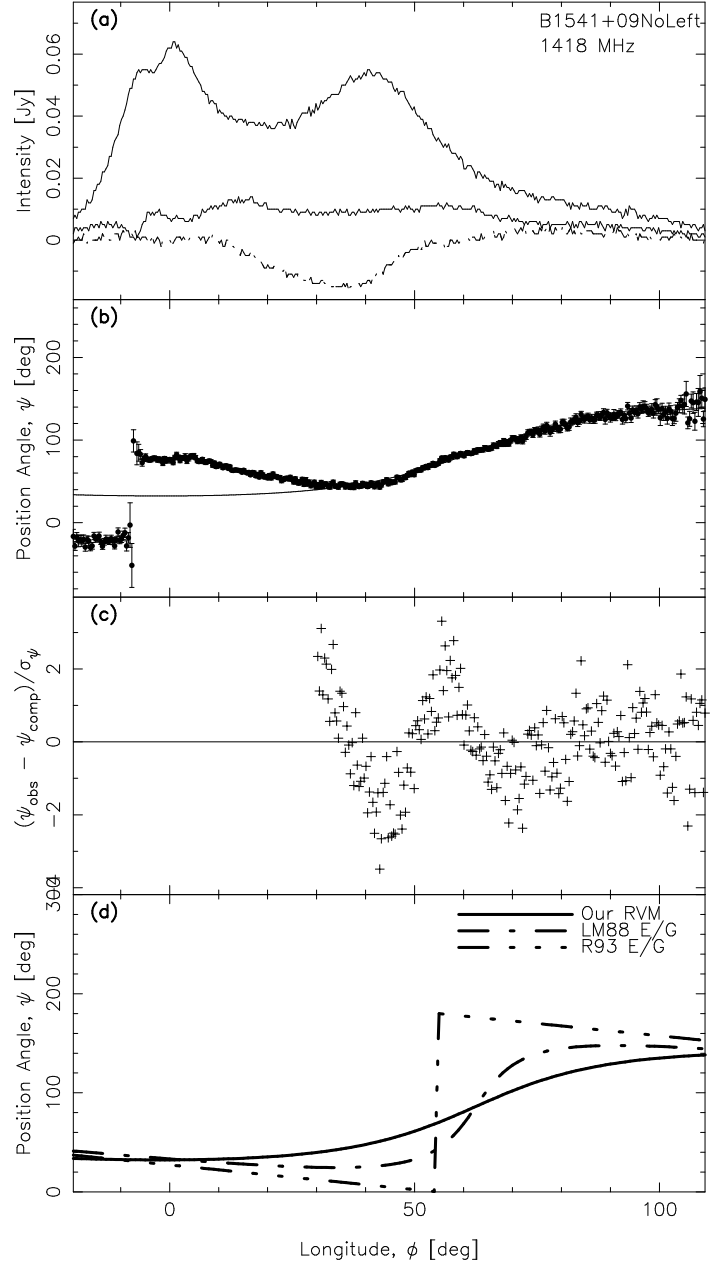}
\caption{Pulsar B1541+09. Narrower longitude fit. For general
explanation of this figure, see Fig. 3 and Fig. 6 captions.
{\em{All}} longitudes below $30\arcdeg$ are unweighted in in this
alternative fit shown in (b) and (c).  See Fig. 11 for a discussion of
the R93 curve in (d).
\label{fig-12}}
\end{figure*}

%fig 13 follows:
\clearpage
\begin{figure*}
\epsscale{0.6}
\plotone{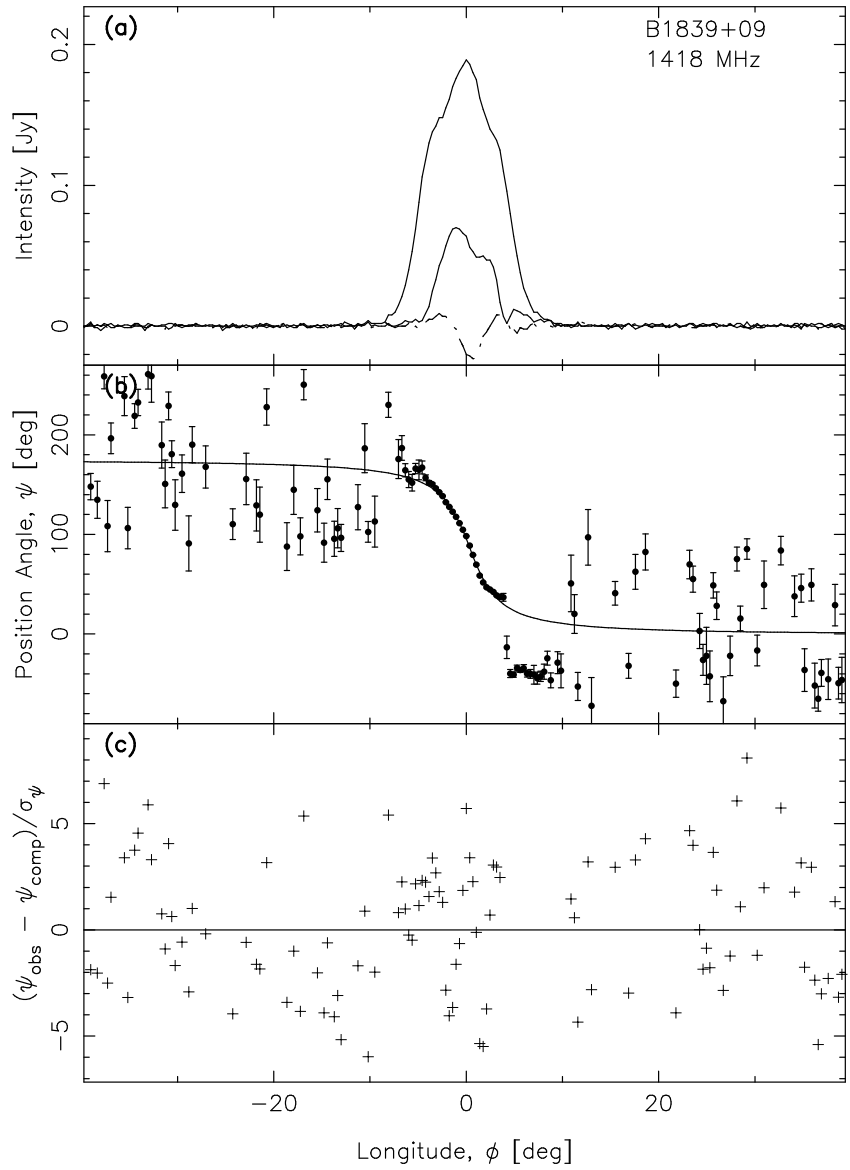}
\caption{Pulsar B1839+09. See Fig. 3 caption for details.  Longitudes between 
$(+4,+10)\arcdeg$  are unweighted in the fit.
\label{fig-13}}
\end{figure*}

%fig 14 follows:
\clearpage
\begin{figure*}
\epsscale{0.6}
\plotone{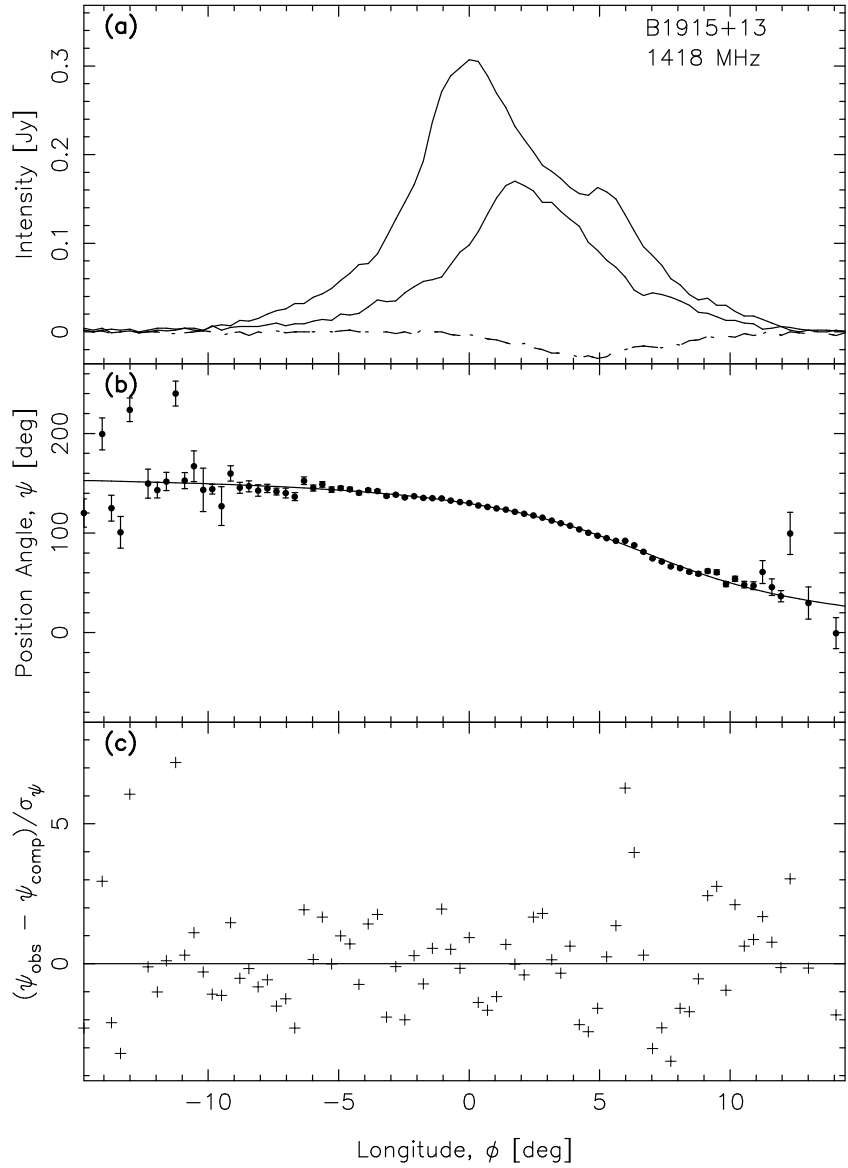}
\caption{Pulsar B1915+13. See Fig. 3 caption for details.
\label{fig-14}}
\end{figure*}

%fig 15 follows:
\clearpage
\begin{figure*}
\epsscale{0.6}
\plotone{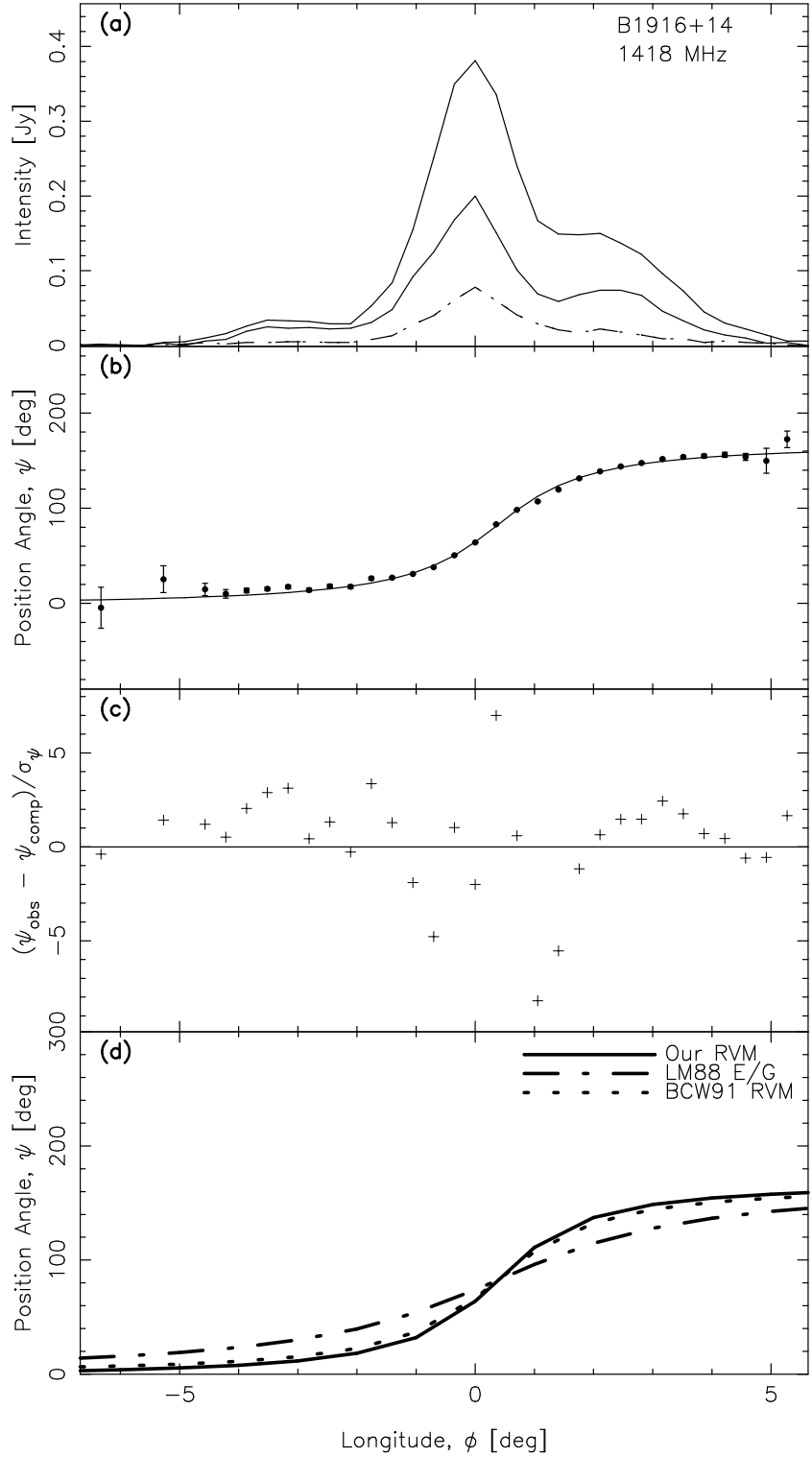}
\caption{Pulsar B1916+14.  For a general explanation of this figure,
see Fig. 3 and Fig. 6 captions.  At the resolution of panel (d), the
R93 E/G curves is indistiguishable from the BCW91 curve, so we do not
plot the R93 E/G curve. 
\label{fig-15}}
\end{figure*}

%fig 16 follows:
\clearpage
\begin{figure*}
\epsscale{0.5}
\plotone{pulsar1929+10Full.080200.ps}
\caption{Pulsar B1929+10. Our preferred full longitude fit.  For
general explanation of this figure, see Fig. 3 and Fig. 6 captions.
Note that a $90\arcdeg$ orthogonal mode switch is inserted in the
longitude range $(-60,-15)\arcdeg$ and that longitudes between
$(-17,+10)\arcdeg$ are unweighted in the fits shown in (b) and (c).
Our preferred RVM curve and most other published RVM fits provide the
best match to the data.  At the resolution of panel (d), the NV82 RVM,
RR97 RVM (their fit), and BCW91 RVM curves are indistiguishable from
our curve, so we do not plot them.  Also, the LM88 RVM (their fit)
curve is indistinguishable from the LM88 E/G (their empirical
calculation) curve, so we do not plot the LM88 RVM curve.
\label{fig-16}}
\end{figure*}

%fig 17 follows:
\clearpage
\begin{figure*}
\epsscale{0.5}
\plotone{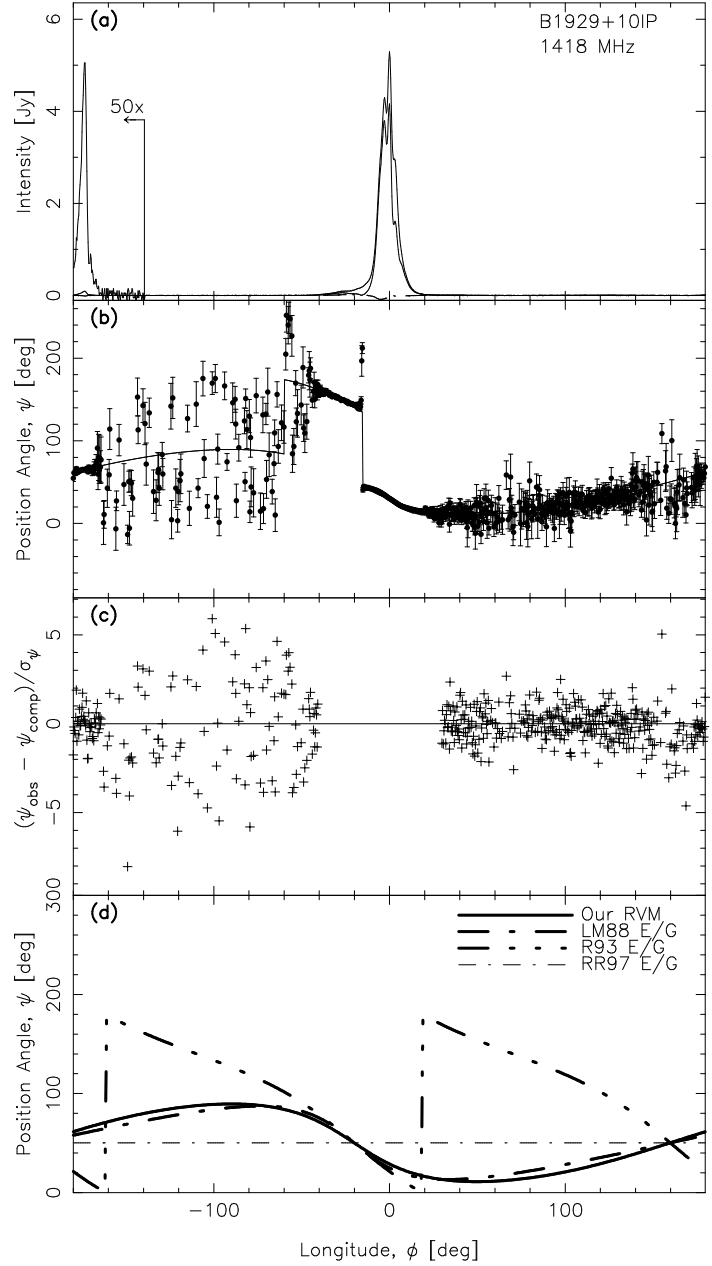}
\caption{Pulsar B1929+10. Entire main pulse region unweighted. For
general explanation of this figure, see Fig. 3 and Fig. 6 captions.
Note that a $90\arcdeg$ orthogonal mode switch is inserted in the
longitude range $(-60,-15)\arcdeg$ and that longitudes between
$(-17,+10)\arcdeg$ are unweighted in the fits shown in (b) and (c).
Our RVM curve and most other published RVM fits provide the best match
to the data.  We omit the same curves as in Fig. 16.
\label{fig-17}}
\end{figure*}
\newpage

\begin{table*}
\caption{Dictionary and Conversion Table from Earlier Work for Geometrical Beam Parameters.
(See Eqs. 1 and 2  and ensuing discussion.)  
\label{tbl-1}}
\scriptsize
\begin{tabular}{|c||c|c|c|c|c|}
%\begin{deluxetable} {|c||c|c|c|c|c|}
%\tablenum{1}
%\tablecolumns{6}
\hline
%\tablehead{ 
Investigators   &   $\psi$  & \multicolumn{2}{c|}{Colatitude of Observable Magnetic Pole, $\alpha$}& \multicolumn{2}{c|}{Impact Parameter of Line of Sight} \\ 
and		&   Convention  & \multicolumn{2}{c|}{w.r.t. ($+\vec{\Omega}$) Spin Axis}	            		   & \multicolumn{2}{c|}{w.r.t. Observable Magnetic Pole} \\
                                                                     \cline{3-6}
 Method         &   Problem?    & Confined   & Relation to             & Symbol in & Relation to \\ 
                &     & to First  & our $\alpha$            & Original  &  our $\beta$ \\ 
                &         & Quadrant? & ($\alpha_{original} \rightarrow$ our $\alpha$) & Paper &  ($Symbol_{original} \rightarrow$ our $\beta$) \\
%}
\hline \hline
%\startdata
Current Work            & no         & no        & $\alpha\rightarrow\alpha$                  & $\beta$              & $\beta\rightarrow\beta$ \\ 
(RVM)                   &            &           &                                            &                      &                         \\ 
\hline
NV82                    & no         & yes       & If $(d\psi/d\phi$ and $\beta_{NV})$        & $\beta_{NV}$         &  If $d\psi/d\phi>0$, then \\  
(RVM)                   &            &           & both have same sign, then                  &                      &  $-|\beta_{NV}|\rightarrow\beta.$ \\
                        &            &           & $\pi-\alpha_{NV}\rightarrow\alpha.$        &                      & If $d\psi/d\phi<0$, then   \\
                        &            &           & Otherwise, $\alpha_{NV}\rightarrow\alpha.$ &                      & $|\beta_{NV}|\rightarrow\beta.$ \\
                        &            &           &                                            &                      &                         \\ 
\hline
LM88                    & n/a        & yes       &  As the sign of  $\beta_{LM}$ was not      & $\beta_{LM}$         & If $d\psi/d\phi>0$, then \\
(E/G)                   &            &         &  published, either $\alpha_{LM}\rightarrow\alpha$ & (magnitude only & $-|\beta_{LM}|\rightarrow\beta.$ \\
                        &            &      &  or  $\pi-\alpha_{LM}\rightarrow\alpha$ are possible. &  was published)& If $d\psi/d\phi < 0$, then \\
                        &            &           &  We selected one based on con-             &                      & $|\beta_{LM}| \rightarrow \beta.$  \\ 
                        &            &           &  sistency with our fits.                   &                      &                                     \\ \hline
BCW91                   & yes        &  no       &   $\pi-\alpha_{BCW}\rightarrow\alpha.$     & $\sigma_{BCW}$       & $\sigma_{BCW} \rightarrow -\beta$ \\ 
(RVM)                   &            &           &                                            &                      &                         \\ 
\hline
R90,93a,b               & n/a        & yes       & If $(d\psi/d\phi$ and $\beta_{R})$         & $\beta_{R}$          & If $d\psi/d\phi>0$, then  \\
 (E/G)                  &            &           & both have same sign, then                  &                      & $-|\beta_{R}|\rightarrow\beta.$ \\
                        &            &           & $\pi-\alpha_{R}\rightarrow\alpha.$         &                      & If $d\psi/d\phi < 0$, then       \\
                        &            &           & Otherwise, $\alpha_{R}\rightarrow\alpha.$  &                      & $|\beta_{R}|\rightarrow\beta.$  \\
\hline
HX97a,b                 & yes        &  no       &   $\pi-|\alpha_{HX}|\rightarrow\alpha.$    & $\sigma_{HX}$        & If $\alpha_{HX} > 0$, then \\ 
(RVM)                   &            &           &                                            &                      & $\sigma_{HX} \rightarrow -\beta.$ \\
                        &            &           &                                            &            & Otherwise, $\sigma_{HX} \rightarrow \beta.^{a}$\\
%  \enddata
\hline \hline
\end{tabular}
\normalsize
\tablecomments{(a) In one case, (PSR B0950+08 at 1.41 GHz), the resulting 
$\zeta$ would then be greater than $\pi.$ Consequently,  $2|\alpha_{HX}|-\sigma_{HX} 
\rightarrow \beta$ must instead be used.}
\end{table*}
%\end{deluxetable}

\newpage
\begin{table*}
\caption{Inner and Outer Line of Sight Trajectories  
\label{tbl-2}}
\normalsize
\begin{tabular}{|c|c|c|c|l|}
\hline
Sign of			     & Sign of 	    &  Sign of Impact Param.   & Colatitude of & Line of Sight Trajectory \\ 
Slope$^{a}$		     & Slope$^{b}$  &  of Line of Sight w.r.t. & Observable Mag- & w.r.t.   \\
$\frac{d\psi}{d\phi}|_{max}$ & $\frac{d\psi'}{d\phi}|_{max}$ & Obs. Mag. Pole, $\beta$ & netic Pole, $\alpha$ & Observable Magnetic Pole \\
                                                                     
\hline \hline
                             &                  &          & $<\pi/2$ & Outer (Equatorward)$^{c}$ \\       
Negative		     & Positive         & Positive &          & \\
                             &                  &          & $>\pi/2$ & Inner (Nearest $[-\vec{\Omega}]$ Spin Poleward)\\
\hline
                             &                  &          & $<\pi/2$ & Inner (Nearest $[+\vec{\Omega}]$ Spin Poleward)\\       
Positive                     & Negative         & Negative &          & \\
                             &                  &          & $>\pi/2$ & Outer (Equatorward)$^{c}$ \\

\hline \hline
\end{tabular}
\normalsize
\tablecomments{(a) The quantity $\frac{d\psi}{d\phi}|_{max}$ is the position angle
sweep rate in the observers' convention (increasing 
counterclockwise on the sky),  while (b) $\frac{d\psi'}{d\phi}|_{max}$ is the sweep rate in the (opposite) RVM 
convention. (c) ``Equatorward'' indicates that the line of sight is {\em{opposite}} the spin pole lying nearest to the 
observable magnetic pole. }
\end{table*}

\newpage
\setlength{\tabcolsep}{0.05in}

%\begin{deluxetable}{|c|c||rrrr|rr|rr|lrr|rr|lrr|}
%\begin{deluxetable}{ccrrrrrrrrlrrrrlrr}
\begin{deluxetable}{ccrrrrrrrrrrrrrrrr}
\tabletypesize{\footnotesize}
\rotate
\tablewidth{0pt}
\tablenum{3}
\tablecolumns{18}
%\tableheadfrac{0.1}
\tablecaption{Geometrical angles $\alpha$ and $\beta$ from our work and 
other investigators.' All angles have been converted to a common
definition using Table~\ref{tbl-1}.  \label{tbl-3}}
\tablehead{Pulsar & R93a,b   &\multicolumn{4}{c}{Our adopted results} &\multicolumn{2}{c}{NV82} & \multicolumn{2}{c}{LM88} &\multicolumn{3}{c}{BCW91} &\multicolumn{2}{c}{R93a,b} & \multicolumn{3}{c}{HX97a,b} \\
&   Class.   &\multicolumn{4}{c}{(1.42 GHz)} &\multicolumn{2}{c}{(0.43 GHz)} & \multicolumn{2}{c}{(0.4 GHz)} &\multicolumn{3}{c}{ } &\multicolumn{2}{c}{1 GHz} & \multicolumn{3}{c}{} \\
& & \multicolumn{4}{c}{(RVM)} & \multicolumn{2}{c}{(RVM)}& \multicolumn{2}{c}{(E/G)} & \multicolumn{3}{c}{(RVM)} & \multicolumn{2}{c}{(E/G)} & \multicolumn{3}{c}{(RVM)} \\ 
%\tablehead{Pulsar & R93a,b   &\multicolumn{4}{c|}{Our adopted results} &\multicolumn{2}{c|}{NV82} & \multicolumn{2}{c|}{LM88} &\multicolumn{3}{c|}{BCW91} &\multicolumn{2}{c|}{R93a,b} & \multicolumn{3}{c|}{HX97a,b} \\
%&   Class.   &\multicolumn{4}{c|}{(1.42 GHz)} &\multicolumn{2}{c|}{(0.43 GHz)} & \multicolumn{2}{c|}{(0.4 GHz)} &\multicolumn{3}{c|}{ } &\multicolumn{2}{c|}{1 GHz} & \multicolumn{3}{c|}{} \\
%& & \multicolumn{4}{c|}{(RVM)} & \multicolumn{2}{c|}{(RVM)}& \multicolumn{2}{c|}{(E/G)} & \multicolumn{3}{c|}{(RVM)} & \multicolumn{2}{c|}{(E/G)} & \multicolumn{3}{c|}{(RVM)} \\ 
& & $\alpha$ & $\beta$ & $\phi_{0}$ & $\chi_{\nu}^{2}$\phn & $\alpha$ & $\beta$\phn & $\alpha$ & $\beta$\phn &  $\nu$ (GHz) &   $\alpha$ & $\beta$\phn & $\alpha$ & $\beta$\phn & $\nu$ (GHz) & $\alpha$ & $\beta$ }
%\hline \hline 
\startdata
0301+19    & $D$ &
     162.4 & 0.96       & -3.71        & 15.6\phn & 
       110 &    3\phn   &
     148.1 &  1.8\phn   & 
      1.42 &   69       &  2.9         &
       150 &  1.7\phn   & 
      4.85 & 85.7       & 13.9 \\ 

           &     &
$\pm11.8$  & $\pm0.63$  & $\pm0.006$   & \phn     &        
  $\pm11$  &        \phn&
           &            &
           & $\pm16$    & $\pm0.3$     &
           &            &
           & $\pm15.0$  & $\pm15.0$  \\ 

           &     &
           &            &              &    \phn  &
           &       \phn &
           &            &
      0.43 &         61 &         2.8  &
           &            &
           &            &          \\

           &     &
           &            &              &    \phn  &
           &       \phn &
           &            &  
           &   $\pm30$  &    $\pm0.4$  &
           &            &
           &            &          \\
\hline
0525+21    & $D$ &
     116.8 & -1.50      & -5.84        &   7.9\phn &
 $\geq160$ &   -0.6\phn & 
     156.8 &   -0.7     &
      1.42 &    134     &  -1.2        &
       159 &   -0.6     &  
      1.41 &   44.2     &  -1.3 \\ 

           &     &
 $\pm4.6$  & $\pm0.08$  &  $\pm0.02$   & \phn      &
           &       \phn &
           &            &
           &   $\pm10$  &   $\pm0.2$   &
           &            &
           &  $\pm60$   &       $\pm0.2$  \\ 

           &     &
           &            &              & \phn      &
           &       \phn &
           &            & 
      0.43 &        140 &         -1.0 &
           &            &
      1.71 &      113.7 & -2.0   \\ 

           &     &
           &            &               & \phn     &
           &       \phn &
           &            &
           &   $\pm9$   &     $\pm0.2$  &
           &            &
           &  $\pm80$   &      $\pm1.0$   \\ 

           &     &
           &            &               &  \phn   &
           &       \phn &
           &            &
           &            &               &
           &            &
      4.85 &      162.2 & -0.6   \\ 

           &     &
           &            &               &   \phn  &
           &       \phn &
           &            &
           &            &               &
           &            &
           &  $\pm70$   & $\pm2.0$   \\ 

\hline
   0656+14 & $T^{a}$ &
       29  &        8.9 &         14.9  & 2.6\phn &
           &       \phn &
       8.2 &        8.2 &
           &            &               &
       30  &            &
           &            &          \\

           &         &
  $\pm23$  &  $\pm6.1$  &    $\pm0.7$   &    \phn &
           &       \phn &
           &            &
           &            &               &
           &            &
           &            &          \\

\hline
   0823+26 & $S_{t}$ &
      98.9 &      -3.03 &         1.33  & 8.6\phn &
        98 &     -4\phn & 
      76.9 &       -1.1 & 
      1.42 &         89 & -3            & 
       84  & $-1.9^{b}$ &
$1.41^{c}$ &       90.2 &     -1.0  \\ 

           &         & 
 $\pm0.7$  & $\pm0.01$ 	&    $\pm0.01$  &    \phn &
           &       \phn &
           &            &
           &  $\pm600$  &   $\pm1$      &
           &            &
$        $ &   $\pm25$  &     $\pm0.2$  \\

    (Main /&         &
           &            &               &    \phn &
           &       \phn &      
           &            &
      0.43 &        101 & -3.2          &
           &            &
$1.71^{c}$ &       97.7 & -1.7   \\

           &         &
           &            &                &   \phn &
           &       \phn &
           &            &
           &    $\pm1$  &      $\pm0.3$  &
           &            &
           &  $\pm80$   &      $\pm1.5$   \\

 Postcursor) &       &
           &            &                &   \phn &
           &       \phn &
           &            &
           &            &                &
           &            &
$4.85^{c}$ &       82.5 & 12.1   \\ 

           &         &
           &            &                &   \phn &
           &       \phn &
           &            &
           &            &                &
           &            &
	   &  $\pm$ 15  & $\pm$ 15   \\ 

 Interpulse &        &
           &            &                &   \phn &
           &       \phn &
 $\pm26.1$ &        5.6 &
           &            &                &
           &            & 
           &            &    \\

\hline
0950+08    & $S_{d}$ & 
     105.4 &       22.1 &           -11.9 & 2.2\phn & 
       170 &      5\phn &
     174.1 &        4.2 &
      1.42 &        174 &             2.5 & 
       168 &        8.5 & 
      1.41 &      179.3 & 0.4     \\

           &         & 
 $\pm0.5$  &  $\pm0.1$  &       $\pm0.2$  &    \phn &
           &       \phn &
           &            &
           &   $\pm30$  &        $\pm15$  &
           &            &
           &   $\pm15$  & $\pm15$       \\

     (Full)&         &
           &            &                 &    \phn &
           &       \phn &
           &            & 
      0.43 &        174 &            2.5  &
           &            &
      4.85 &      153.1 & 4.1   \\  

           &         &
           &            &                 &    \phn &
           &       \phn &
           &            &       
           &   $\pm90$  &        $\pm40$  &
           &            &
           &  $\pm90$   & $\pm5$    \\  
\hline
   1541+09 &   $T$  & 
     131.0 &     -20.23 &          62.5   & 1.5\phn  &
           &       \phn & 
     174.2 &       -0.8 & 
           &            &                 & 
       175 &        0.0 & 
           &            & \\

           &        &
$\pm5.67$  & $\pm2.26$  &     $\pm0.37$   &     \phn &
           &       \phn &
           &            &
           &            &                 &
           &            &
           &            & \\
\hline
   1839+09 &   $T$  & 
      86.1 &        2.3 &           0.34  & 10.3\phn &
           &       \phn & 
        90 &        2.9 &
      1.42 &        146 &               1 & 
        97 &  $1.4^{b}$ & 
           &            & \\ 

           &        &
$\pm11.4$  & $\pm0.04$  &     $\pm0.02$   &     \phn &
           &            &
           &            &     
           &  $\pm400$  &         $\pm8$  &
           &            &
           &            & \\ 

\hline
   1915+13 & $S_{t}$& 
        73 &        5.4 &             6.8 &   4.2\phn &
           &            &
           &            &
      1.42 &         86 &             5.4 &
        68 &        6.6 &
      4.85 &       89.3 &  9.3  \\ 

           &            &      $\pm19.4$  &  $\pm0.5$  &
$\pm0.1$   &       \phn & 
           &            & 
           &            &
           &   $\pm24$  &       $\pm0.2$  &
           &            &
           & $\pm12.5$ 	& $\pm15$  \\ 
\hline
   1916+14 & $T(?)$ &
     118.0 &    $-$1.0  &          0.38   &    8.9\phn &
           &            & 
     124.5 &       -2.1 &
      1.42 &        167 &           -0.3  &
       101 &       -1.3 & 
           &            & \\ 

           &        &
$\pm33.4$  & $\pm0.33$  &      $\pm0.01$   &      \phn &
           &            &
           &            &
           &  $\pm100$  &        $\pm0.2$  &
           &            & 
           &            & \\ 
\hline
$1929+10^{d}$& $T(?) /$ & 
     35.97 &      25.55 &           -19.62 &   2.45\phn &
        35 &         23 & 
         6 &          4 &
      1.42 &         27 &               16 &
   $90^{f}$&       41.8 & 
           &            & \\

           &        &
$\pm0.95$  & $\pm0.87$  &        $\pm0.2$  &      \phn &
           &            &
           &            &
           &    $\pm4$  &          $\pm3$  &
           &            & 
           &            & \\

           & $cT$   &
           &            &                  &     \phn &
           &            &
           &            &
 $1.42^{e}$&        150 &               -3 &
   $18^{f}$&       11.6 & 
           &            & \\ 

           &        &
           &            &                 &      \phn &
           &            &
           &            &
           &   $\pm10$  &         $\pm2$  &
           &            & 
           &            & \\ 

           &        &
           &            &                 &     \phn &
           &            &
           &            & 
      0.43 &         25 &              16 &
           &            & 
           &            & \\ 

           &        &
           &            &                 &          &
           &            &
           &            &
           &   $\pm2$   &         $\pm2$  &
           &            &
           &            & \\ 

Interpulse &        &
           &            &                 &     \phn &
           &            &
           &            &
           &            &                 & 
        88 &            & 
           &            & 
\enddata
%\hline \hline
\tablecomments{(a) W99 suggest $S_{t}$; (b) Here we choose opposite sign as R93a,b, as required by Eq. (5); (c) Main pulse only; 
(d) See text for additional RVM fits by LM88,P90, and  RR97; (e)  Alternate BCW91 fit; (f) First Rankin fit is under assumption 
of $T$ classification; second is for $cT$.}
\end{deluxetable}

\newpage
%\setlength{\tabcolsep}{0.04in}

%\begin{deluxetable}{|c|c|c||rrrr|}
\begin{deluxetable}{cccrrrr}
\tabletypesize{\normalsize}
\tablewidth{0pt}
\tablenum{4}
\tablecolumns{7}
\tablecaption{Multiple convergent fits to pulsars with multiple components.  \label{tbl-4}}
\tablehead
{Pulsar & Components fitted 	& Longitude	& \multicolumn{4}{c}{Our  results}  \\
	& 			& limits$^a$	& $\alpha$ 	& $\beta$	& $\phi_{0}$ 	& $\chi_{\nu}^{2}$  	}
%\hline \hline 
\startdata
0823+26	& full 			& $\pm$180	& 98.9  	& -3.03 	& 1.33 		& 8.6 			\\
        &       		&		& $\pm$ 0.7	&   $\pm$ 0.01 	& $\pm$ 0.01 	&			\\

	& 			&		&		&		&		&			\\
	& main / postcursor	&$-20$ to $45$	& 85.8		& -3.08		& 1.21		& 8.6			\\
	& 			&		& $\pm$ 3.2	&   $\pm$ 0.01	& $\pm$ 0.02 	&			\\
\hline

0950+08 & full 			& $\pm180$	&  105.4     	& 22.1       	&     -11.9 	& 2.2      		\\
	&       		&		& $\pm$ 0.5	& $\pm$ 0.1 	& $\pm$ 0.2 	&			\\

	& 			&		&		&		&		&			\\
	& main pulse		&$-90$ to $+90$ & 126.2		& 19.6		& -10.8	 	& 1.3			\\
	& 			&		& $\pm$ 4.3     &$\pm$1.0	 &$\pm$ 0.3	&			\\

	& 			&		&		&		&		&			\\
	& interpulse 		&$+90$ to $-90$	& 109.6		& 13.3 		& -37.6	 	& 2.3			\\
	& 			&		&$\pm$ 1.4	& $\pm$	0.9	&$\pm$5.3	&			\\
\hline
1929+10 & full 			& $\pm180$	& 35.97 	& 25.55 	& -19.62 	& 2.45			\\
	&                  	&		& $\pm$ 0.95 	& $\pm$ 0.87 	& $\pm$ 0.2  	&			\\

	& 			&		&		&		&		&			\\
	& interpulse 		&$-180$ to $-40$& 34.3		& 28.8
 & -20.2	& 2.5		\\
	& 			&$35$ to $180$	&$\pm$ 0.86      &$\pm$0.23	 &$\pm$ 1.2		&			\\
\enddata
%\hline \hline
\tablecomments{(a) See text and figure captions to find ranges of unweighted longitudes within these limits.}
\end{deluxetable}

\end{document}